\documentstyle[preprint,aps,floats,epsfig]{revtex}

\newcommand{\be}{\begin{equation}}
\newcommand{\ee}{\end{equation}}
\newcommand{\bea}{\begin{eqnarray}}
\newcommand{\eea}{\end{eqnarray}}
\newcommand{\bml}{\begin{mathletters}}
\newcommand{\eml}{\end{mathletters}}

\begin{document}

\titlepage

\preprint{hep-th/0112085}

\bigskip

\draft

\tighten

\title{Global Black Branes (Extended Global Defects Surrounded by Horizons),
Brane Worlds and the Cosmological Constant}

\author{ {Sei-Hoon Moon}
         \footnote{E-mail: \texttt{ jeollo$@$phya.yonsei.ac.kr}}}

\address{Institute of Physics and Applied Physics, Yonsei University,
Seoul 120-749, Korea }

\date{\today}

\setlength{\footnotesep}{0.5\footnotesep}

\maketitle

\begin{abstract}
We study global defects coupled to higher-dimensional gravity with a
negative cosmological constant. This paper is mainly devoted to studying
global black brane solutions which are extended global defects surrounded
by horizons. We find series solutions in a few separated regions and confirm
numerically that they can be mutually connected. When the world volume
of the brane is Ricci-flat, the brane is surrounded by a degenerated
horizon, while it is surrounded by two horizons when the world volume
has a positive constant curvature. Each solution corresponds to an
extremal and a non-extremal state, respectively. Their causal structures
resemble those of the Reissner-Nordstr\"{o}m black holes in anti-de
Sitter spacetime. However, the non-extremal black brane is not a static
object, but an inflating brane. In addition, we briefly discuss a brane
world model in the context of the global black branes. We comment on a
few thermodynamic properties of the global black branes, and discuss a
decrease of the cosmological constant on the brane world through the
thermodynamic instability of the non-extremal global black brane.
\end{abstract}

\vspace{10mm}








\newpage

\section{Introduction}\label{secInt}
~\\
Global defects are topologically stable scalar field configurations with
nontrivial homotopy for some internal symmetry manifolds \cite{VS}.
Usually, this kind of topological defect arises in a theory with $d$
scalar fields and a potential which has a vacuum manifold with the
topology of a $(d-1)$-sphere. The simplest examples are domain
walls \cite{vilenkin,widrow,gibbons} when the vacuum manifold has a
discrete symmetry, global strings \cite{gss,CK1} when $d=2$, and global
monopoles \cite{BV} when $d=3$, in four spacetime dimensions. While
the global monopole has a well-defined static metric, the domain
wall and the global string do not. The domain wall spacetime is
non-static and the metric in the plane of the domain wall is that
of a $(2+1)$-dimensional de Sitter space. The spacetime of the
global string has a curvature singularity at a finite distance
from the string core \cite{gss}. Its metric was found by Cohen
and Kaplan \cite{CK1}. However, Gregory showed that time dependence
could remove the singular nature of the global string
spacetime \cite{nsgs}. While these analyses were purely within the
context of Einstein gravity in four dimensions (in the absence of
a cosmological constant), the global string spacetime is always
regular in the presence of a negative cosmological
constant \cite{Gre}, no matter how small cosmological constant may
be. Because the negative cosmological constant could do a role to avoid
the singularity rounding off the spacetime before it terminates at
the singularity.

Global defects were first considered  in four spacetime
dimensions, but recently they have been studied in higher
dimensions in the context of the brane world scenarios, such as
the large extra dimension scenario (ADD) \cite{ADD} and the
Randall-Sundrum (RS) scenario \cite{RS1,RS2}. According to the brane
world picture the three space dimensional world that we appear to
be living in is a brane that is embedded in a higher-dimensional
world. The brane world scenarios are really attractive in that the
Newtonian gravity is recovered on the brane despite the gravity is
fundamentally higher-dimensional one, even with large (infinite)
extra dimensions. This approach provides a plausible explanation
of the hierarchy between the gravitational and the electroweak
mass scales, opening the possibility for construction of new
classes of models for the unified theory. These theories also open
new approaches to the cosmological constant problem, because if
our four-dimensional world is embedded in a higher-dimensional
spacetime, the effect of non-zero vacuum energy could affect only
the curvature in the extra dimensions allowing for a flat
four-dimensional world as first proposed in Ref.
\cite{rubakov,ADKS,KMS,KT,tetra,KKL}.

There have been various attempts to realize the brane world scenario
in the context of higher-dimensional gravitating global defects in
Refs. \cite{CK2,Gre,Vil,shap1,shap,Oda,KMR,inyong}. Cohen and
Kaplan \cite{CK2} have considered a brane world, which may be
viewed as a global string in two extra dimensions. Its spacetime
exhibits a curvature singularity at a finite distance from the
string core, similar to the case in four dimensions. They argued
that, imposing a unitary boundary condition, the location of the
singularity may play a role of the boundary of the extra
dimensions, which form a finite but non-compact space of
exponentially large proper size. The phenomenology resembles the
ADD scenario, rather than that of the RS scenario. Gregory \cite{Gre}
has shown that a non-singular global string-like solution with the
required properties for the RS-type scenario exists in the presence of
a negative bulk cosmological constant. Olasagasti and
Vilenkin \cite{Vil} explored a more general case of a brane
carrying a global charge in a higher-dimensional spacetime with a
nonzero cosmological constant. In particular, with a negative
cosmological constant there exist solutions of which the geometry
of extra dimensions asymptotes to a cylinder with cross section
being a $(d-1)$-sphere of a fixed radius (cigar-like geometry).
The cigar-like solutions with an exponentially decaying warp
factor is of interest, since they have features needed for the RS-type
scenario.

On the other hand, in Ref. \cite{KMR} the authors investigated the
spacetime of higher-dimensional global defects in the presence of
a negative cosmological constant adopting Schwarzschild-type
metric ans\"{a}tz. They found extremal black hole-like $p$ space
dimensional defects (global black $p$-branes), which are
Ricci-flat branes surrounded by a degenerated horizon. They showed
that the near horizon geometry of the extremal global black branes
coincides with the cigar-like geometry with an exponentially
damping warp factor discovered in Refs. \cite{Gre,Vil}. The
interior region inside the horizon of the global black $p$-brane
possesses all of features needed for a brane world with the Newtonian
gravity. In the picture of Ref. \cite{KMR} the size of the horizon can be
interpreted as the compactification size of $d$ extra dimensions,
even though the interior region inside the horizon infinitely
extends and asymptotes to $AdS_{p+2}\times S^{d-1}$. Therefore,
the large mass hierarchy is translated into the large size of the
horizon, which could be supported by the large magnitude of charge
carried by the black branes. In this picture, the Hawking
radiation could be a possible mechanism for the resolution of
various problems associated with brane world scenarios, such as
the observed flatness and the approximate Lorentz invariance of
our world. The bending of the brane and the bulk curvature
that violate the $SO(3,1)$ isometry on our brane \cite{kolb,poritz}
would correspond to excitations upon the extremal state, and the
excited states that are non-extremal black branes would evolve
into an extremal state through the Hawking radiation process.

The nonzero vacuum energy density (equivalently, the cosmological
constant) on the brane could also be an excitation upon the extremal
state, which would be diluted via the Hawking process if the
excited state correspond to a non-extremal black brane. However,
even though the state excited by the vacuum energy density is a
non-extremal state, it will be somehow different from static
non-extremal black branes treated usually in the literatures
before in the context of supergravity or string theories \cite{HS},
because with nonzero vacuum energy density the branes will inflate
and the corresponding spacetime will not be static. Therefore, their
thermodynamic properties will also be different from those of the
static ones. In this respect, we may need to find the corresponding
black brane solutions and to examine their thermodynamic properties,
to see whether a certain thermodynamic process can be a dynamical
mechanism thinning out the vacuum energy density on the brane.

This paper will be mainly devoted to finding global black brane
solutions which are extended global defects surrounded by horizons
and studying their spacetime structures. However, we will also
focus on a few aspect of a brane world based on the global black
branes and the cosmological constant problem.

This paper is organized as follows. In Sec. II, we derive the
equations of motion corresponding to the Schwarzschild-type metric
ans\"{a}tz in a scalar theory with global internal symmetry
coupled to higher-dimensional gravity with a negative cosmological
constant. In Sec. III, we explore black hole-like $p$ space
dimensional defect (black $p$-brane) solutions, which shall be referred
in what follows as \lq\lq global black $p$-brane" or \lq\lq global
black brane". We find series solutions in a few separated regions
and confirm numerically that they can be mutually connected. When
the world volume of the branes is Ricci-flat, the branes are
surrounded by a degenerated horizon, while when the world volume
is not Ricci-flat but of a positive constant curvature, they are
surrounded by two horizons.
They correspond to extremal black branes and non-extremal ones,
respectively. In Sec. IV, we discuss the spacetime structures of
the global black branes. Their causal structures resemble those of
the Reissner-Nordstr\"{o}m black holes in anti-de Sitter
spacetime, if we consider the transverse slice to the brane. The
near horizon geometries are an infinitely long anti-de Sitter
throat with the topology of $AdS_{p+2}\times S^{d-1}$ for the
extremal black branes, and the spacetime of an inflating $p$ space
dimensional domain wall times $(d-1)$-dimensional sphere for non-extremal
ones. In Sec. V, we discuss a few issues associated to the brane world
scenarios. We briefly comment a few thermodynamic
properties of the global black branes, and discuss a decrease of the
cosmological constant on the brane world through the thermodynamic
instability of the non-extremal global black brane. In the last section,
we end with some conclusions.

\section{The Equations of Motion}\label{secEoM}
~\\
In this section we derive field equations appropriate to isolated
gravitating global defects in higher-dimensional spacetime with
a negative cosmological constant. Higher-dimensional gravitating
global defects are described as topologically stable scalar field
configurations of a field theory with a spontaneously broken
continuous global symmetry coupled to $D$-dimensional gravity.
We consider a scalar theory with global $O(d)$ symmetry coupled to
gravity, of which potential has minimum on the $(d-1)$-sphere of
radius $v^2$:
\begin{eqnarray}
S&=&S_{{\rm gravity}}+S_{{\rm scalar}}  \nonumber\\
S_{{\rm gravity}}&=& \frac{M^{D-2}_*}{16\pi}\int dx^D\sqrt{g_{D}}~
(R-2\Lambda) \label{action1}\\
S_{{\rm scalar}}&=&\int dx^D\sqrt{g_{D}}\left[
-\frac12 \nabla_M\phi^a\nabla^M\phi^a -
\frac{\lambda}{4} (\phi^b\phi^b -v^2)^2 \right] \label{action},
\end{eqnarray}
where $M_*$ is the fundamental scale of the higher-dimensional
gravity theory. We use a mostly plus signature. Note that $\phi^a$
and $v$ have a mass dimension of $(p+d-1)/2$ and the negative bulk
cosmological constant ($\Lambda<0$) has a mass dimension of 2.
This theory admits $p$ space dimensional topological solitons,
where $p+1=D-d$. We shall refer them as `global $p$-brane' in what
follows. $M, N, ...$ denote $D$-dimensional spacetime indices and
$a, b, ...$ do $d$-dimensional internal space indices and run
on $1, ..., d$. We shall use the notation $\{x^\mu\}$ with
$\mu=0,...,p$ for the coordinates on the $p$-brane world volume.
$\{y^a\}$ denote the transverse coordinates to the $p$-brane and
are related to the spherical coordinates of the transverse space
by the usual relations:
$y^a=\{r\cos\theta_1,\cdots\cdots,r\sin\theta_1 \cdots
\cos\theta_{d-1}, r\sin\theta_1\cdots\sin\theta_{d-1}\}$ with
$r^2=y^a y^a$. The corresponding Einstein equations and scalar
field equation are
\begin{eqnarray}
R_M{}^N=\frac{8\pi}{M_*^{D-2}}{\cal T}_M{}^N+\delta_M{}^N\frac{2}
{D-2}\Lambda,\label{ein}\\
\nabla_M\nabla^M\phi^a-\lambda(\phi^b\phi^b-v^2)\phi^a=0,\label{scala}
\end{eqnarray}
where
\begin{equation}
{\cal T}_M{}^N\equiv \nabla_M\phi^a\nabla^N\phi^a
+\delta_M{}^N\frac{\lambda}{2(D-2)}(\phi^b\phi^b-v^2)^2. \label{stensor}
\end{equation}

The global defects are defined by a topologically nontrivial mapping of the
vacuum manifold $O(d)/O(d-1)$ to the boundary of transverse dimensions
$S^{d-1}$. We take ans\"{a}tz for the scalar field configuration as:
\begin{equation}\label{hedg}
\phi^a=vX~y^a,
\end{equation}
where the function $X$ could be a function of both $t$ and $r$ in general.
The defect solution should have $X(r=0)=0$ at the center of the defect and
approach the radial `hedgehog' configuration outside the core.

In this paper, we consider a particular case that the
energy-momentum tensor of the defects is time independent. As a
result, no gravitational and particle radiation exists. The
spacetime section transverse to the defects is static, and the
whole spacetime including longitudinal directions is allowed to be
static or stationary as we will see later.
The spacetime of the extended global defects have the
topology of $R^{p+2}\times S^{d-1}$. For black branes, the
spacetime transverse to the brane will be a generalization of the
$4$-dimensional spherically symmetric solutions, {\it e.g.}
the Schwarzschild or the Reissner-Nordstr\"{o}m black hole solutions.
To obtain these solutions of the Einstein equations we will assume
that outside any horizon there is a Killing vector which is
timelike and that surfaces of constant $t$ are formed from
concentric $(d-1)$-spheres times $p$-dimensional space. We can
then quite generally adopt following Schwarzschild-type metric
ans\"{a}tz:
\begin{equation}\label{bbm}
ds^2=e^{2A(r)}B(r)\hat{g}_{\mu\nu}(x)dx^\mu dx^\nu+\frac{dr^2}{B(r)}
+r^2d\Omega_{d-1}^2,
\end{equation}
where $\hat{g}_{\mu\nu}(x)$ is a metric on $(p+1)$-dimensional
brane world volume which is not necessarily static,
$d\Omega_{d-1}^2$ is the line element on the unit $(d-1)$-sphere,
and $A(r)$ and $B(r)$ are functions of $r$ only. With this metric
ans\"{a}tz, the defect solutions with unit winding number
correspond to the field configurations
\begin{equation}\label{bbhh}
\phi^a=vf(r)\frac{y^a}{r}.
\end{equation}
The defect solutions have $f(r)=0$ at the center and approach
$f(r)\simeq 1$ outside core. The energy-momentum tensor for the
field configuration (\ref{bbhh}) is then given by
\begin{eqnarray}
\hat{\cal T}_t{}^t&=&\hat{\cal T}_{x_i}{}^{x_i}
    =\frac{(1-f^2)^2}{2(p+d-1)} ,\\
\hat{\cal T}_r{}^r&=&Bf'^2+\frac{(1-f^2)^2}{2(p+d-1)} ,\\
\hat{\cal T}_{\theta_a}^{\theta_a}&=&\frac{f^2}{r^2}
+\frac{(1-f^2)^2}{2(p+d-1)},
\end{eqnarray}
where $\hat{\cal T}^{M}_{\; N} ={\cal T}^{M}_{\; N}/\lambda v^{4}$.

Since the metric (\ref{bbm}) is a special case of the more general
class of metric,
\begin{equation}\label{gem}
ds^2=F(y^a)^2d\hat{s}^2+d\tilde{s}_d^2,
\end{equation}
where $d\tilde{s}_d^2$ is the metric of the slice in the
transverse directions and $d\hat{s}^2$ is the world volume metric,
the Ricci tensor splits in the following way:
\begin{eqnarray}
R_{\mu\nu}&=&\hat{R}_{\mu\nu}-\hat{g}_{\mu\nu}[F\tilde{\nabla}^2F
+p(\tilde{\nabla}F)^2]  \label{rtm}\\
R_{ab}&=&\tilde{R}_{ab}-(p+1)\frac{\tilde{\nabla}_a\tilde{\nabla}_b F}{F}.
\end{eqnarray}
Since ${\cal T}_{\mu\nu}\propto \hat{g}_{\mu\nu}$, through
Einstein's equations we have that $R_{\mu\nu}\propto
\hat{g}_{\mu\nu}$ and finally, from Eq. (\ref{rtm}) above, that
$\hat{R}_{\mu\nu}\propto \hat{g}_{\mu\nu}$. That is, $\hat{R}$,
the curvature associated with the metric $\hat{g}_{\mu\nu}$, must
be constant\cite{Vil}. Therefore, without losing generality we can assume
that the spatial part of the metric intrinsic to the brane is
homogeneous and isotropic, and the directions parallel to the
brane are boost invariant in the strong sense. Einstein's field
equations then reduce to
\begin{eqnarray}
\frac{1}{p+1}\frac{\hat{R}}{e^{2A}B}-(p+1)BA'^2-\frac{2p+3}{2}B'A'
-\frac{p}{4}\frac{B'^2}{B}-BA''-\frac12B'' \nonumber\\
-\frac{d-1}{r}\left(BA'+\frac12B'\right)
=8\pi G_D \hat{{\cal T}}_t{}^t+\frac{2}{p+d-1}\Lambda, \label{ein1}
\end{eqnarray}
\begin{equation}
-(p+1)\left(BA'^2+\frac32B'A'+BA''+\frac12B''\right)
-\frac{d-1}{2}\frac{B'}{r}
=8\pi G_D \hat{{\cal T}}_r{}^r+\frac{2}{p+d-1}\Lambda,\label{ein2}
\end{equation}
\begin{equation}
-(p+1)\frac{BA'}{r}-\frac{p+2}{2}\frac{B'}{r}-(d-2)\frac{B-1}{r^2}
=8\pi G_D \hat{{\cal T}}_{\theta_a}{}^{\theta_a}
+\frac{2}{p+d-1}\Lambda\label{ein3},
\end{equation}
supplemented by the equation for the metric on the $p$-brane,
\begin{equation}
\hat{R}_{\mu\nu}=\frac{\hat{R}}{p+1}\hat{g}_{\mu\nu}. \label{ein4}
\end{equation}
It can be easily shown that only two of the three equations
(\ref{ein1})-(\ref{ein3}) are independent.
The scalar field equation is written down as
\begin{equation}\label{sfe}
f''+\left[\frac{d-1}{r}+\left(1+\frac{p}{2}\right)\frac{B'}{B}+(1+p)A'
\right]f'
-\frac{1}{B}\left[\frac{(d-1)}{r^2}+(f^2-1)\right]f=0.
\end{equation}
Here we have defined a dimensionless Newtonian constant $G_D\equiv
v^{2}/M_{\ast}^{D-2}$ and we have rescaled the coordinates and
redefined the cosmological constant as
\begin{equation}\label{res}
\sqrt{\lambda} v x_M\rightarrow x_M,~~{\rm and}
~~~\Lambda/\lambda v^{2}\rightarrow\Lambda.
\end{equation}

\section{Solutions}\label{secSols}
~\\
We may be able to find several classes of solutions to the set of field
equations (\ref{ein1})-(\ref{sfe}). However, we will restrict our discussion to
only black hole-like solutions in this paper. First, we find extremal black
brane solutions which are flat branes surrounded by a degenerated horizon.
Next we shall discuss about non-extremal black brane solutions which are bent
branes surrounded by two horizons.

For the simplest case (of $p=0$ and $d=2$) which corresponds to a
vortex in $(2+1)$-dimensional spacetimes, the global vortex
solutions were thoroughly analyzed for the global $U(1)$ scalar
theory in Ref.\cite{KKK} and for the $O(3)$ nonlinear $\sigma$
model in Ref.\cite{KM} in the presence of a negative cosmological
constant. For the global $U(1)$ theory, there are three types of
solutions depending on the relative scales of parameters of the
theory: the cosmological constant, the Planck scale and the
symmetry breaking scale. They are regular vortex, extremal and
non-extremal Reissner-Nordstr\"{o}m-type black holes carrying
topological charges. Their spacetimes are perfectly regular and do
not involve any physical curvature singularity. When the magnitude
of a negative cosmological constant is larger than a critical
value at a given symmetry breaking scale, the spacetime formed by
a isolated vortex is regular hyperbola with a deficit angle.
However, it becomes a charged black hole when the magnitude of the
cosmological constant is less than the critical value. For the
$O(3)$ nonlinear $\sigma$ model, the usual regular topological
lump solution cannot form a black hole even though the scale of
symmetry breaking is increased. There exist non-topological
solitons of half integral winding in the given model, and the
corresponding spacetimes involve charged BTZ black holes.

When $p\geq1$ and $d\geq2$, the equations of motion, however, become much more
nonlinear as can be seen from Eqs. (\ref{ein1})-(\ref{sfe}). Due to this high
nonlinearity of Eqs. (\ref{ein1})-(\ref{sfe}), the spacetime around the global
$p$-brane may develop a curvature singularity in an analogous fashion to the
four-dimensional global string metric (with vanishing cosmological constant)
\cite{gss,CK1}. In other side, we may also expect that the bulk negative
cosmological constant could do a role to avoid the singularity
rounding off the spacetime before it terminates at the singularity \cite{Gre}.
As discussed in Ref. \cite{Vil}, the spacetime structure of global extended
defects are depends on the number of extra dimensions $d$, the bulk
cosmological constant $\Lambda$, and the $p$-brane world volume
curvature $\hat{R}$.
For a negative cosmological constant. When $\hat{R}=0$, every spacetime found
in Ref. \cite{Vil} are regular without any genuine singularity.
When $\hat{R}\neq0$, some of the spacetimes has singularity at a finite proper
distance from the core.

We will see that, for black hole-like solutions, cases
$\hat{R}=0$ correspond to extremal black $p$-branes, while cases $\hat{R}\neq0$
to the non-extremal states. The black brane solutions are perfectly
regular ones. Moreover, we will show that some of the independent solutions
found in Ref. \cite{Vil} actually describe different regions of the spacetime
of a global black brane.

\subsection{Extreme black branes: flat branes surrounded by a
degenerated horizon}\label{secSolsA}
~\\
Now we will analyze the equations derived in the previous section for the
isolated global defect spacetime and find solutions of which world volume is
Ricci-flat, i.e., $\hat{R}=0$. We will see that such branes correspond to
extremal black branes, if it has a horizon.
Since it seems almost impossible to find out exact analytic solutions of
equations of motion Eqs. (\ref{ein1})-(\ref{sfe}), we will just try to examine
the behavior of the global $p$-brane spacetime at a few separated regions.
Outside the core, the energy-momentum tensor is approximated into
\begin{eqnarray}
\hat{\cal T}_t{}^t=\hat{\cal T}_{x_i}{}^{x_i}\simeq\hat{\cal T}_r{}^r\simeq0
~~~{\rm and}~~~\hat{\cal T}_{\theta_a}^{\theta_a}\simeq\frac{1}{r^2},
\end{eqnarray}
with $f(r)\simeq1$. Therefore, outside the core the behavior of the solutions
is determined solely by the relative magnitude of the scalar field
energy density ($8\pi G_D/r^2$) to the cosmological constant ($|\Lambda|$)
through the equation (\ref{ein3}).

Near the origin where the field energy density dominates over the cosmological
constant, the series solution up to the leading term is as follows:
\begin{eqnarray}
f(r)&\approx& f_0~ r, \label{frc} \\
A(r)&\approx& A_0 + A_2~r^2,  \label{arc}\\
B(r)&\approx& 1 + B_2~r^2,   \label{brc}
\end{eqnarray}
where $f_0$ is a shooting parameter to be determined by the proper behavior of
the fields at the asymptotic region, and $A_0$ can be consistently fixed as
$A_0=0$ so that the spacetime is locally Minkowski at the origin.
The coefficients $A_2$ and $B_2$ are given as follows
\begin{eqnarray}\label{a2b2}
A_2&\equiv& \left[\frac{4\pi G_D}{d-1}\left(f_0^2-\frac{p}{2d(p+d-1)}\right)
            +\frac{p|\Lambda|}{d(d-1)(p+d-1)}\right],\label{a2} \\
                                                   \nonumber\\
B_2&\equiv&-\left[\frac{8\pi G_D}{d-1}\left(f_0^2-\frac{p-d+1}{2d(p+d-1)}
            \right) +\frac{2(p-d+1)|\Lambda|}{d(d-1)(p+d-1)}\right]\label{b2}.
\end{eqnarray}
While the metric function $A(r)$ increases, $B(r)$ decreases near the origin
for generic values of $f_0$.
With this solution the metric Eq. (\ref{bbm}) near the core takes the form of
\begin{equation}\label{cores}
ds^2\approx\left[1-\frac{4\pi G_D-2|\Lambda|}{d(p+d-1)} r^2\right]
\bar{g}_{\mu\nu}(x)dx^\mu dx^\nu +\frac{dr^2}{1+B_2r^2}+r^2d\Omega_{d-1}^2,
\end{equation}
where $\bar{g}_{\mu\nu}(x)$ is a general Ricci-flat metric on the brane, which
satisfies $(p+1)$-dimensional vacuum Einstein equations
$\bar{R}_{\mu\nu}(\bar{g})=0$.
This shows that the metric component $g_{00}$ of the spacetime decrease
in the transverse directions near the origin when $2\pi G_D>|\Lambda|$,
regardless value of $f_0$.
This fact implies a possibility that the $(00)$-th component of the metric
could vanish at a finite coordinate distance and of the formation of black
hole-like defects, when $2\pi G_D>|\Lambda|$.

In the far region from the core where the cosmological constant dominates over
the scalar fields energy density, the metric functions behave up to the leading
terms for sufficiently large $r$ as follows:
\begin{eqnarray}
f(r)&\approx&1, \label{fra}\\
A(r)&\approx& A_{\infty}+\frac{(p+d)(p+d-1)(8\pi G_D-d+2)}{8(p+1)|\Lambda|}
~\frac{1}{r^2}, \label{ara}\\
B(r)&\approx& \frac{2|\Lambda|}{(p+d)(p+d-1)} ~r^2 \label{bra},
\end{eqnarray}
where $A_\infty$ is a constant to be determined by matching with series
solutions in other regions. The metric (\ref{bbm}) then is
\begin{equation}\label{bbas}
ds^2\approx B_{\infty}r^2~\bar{g}_{\mu\nu}(x)dx^\mu dx^\nu
+\frac{dr^2}{B_{\infty}r^2}+r^2d\Omega_{d-1}^2,
\end{equation}
where $\exp(2A_{\infty})$ has been absorbed into the longitudinal coordinates
$x_\mu$ and $B_{\infty}\equiv 2|\Lambda|/(p+d)(p+d-1)$.
This asymptotes to $D$-dimensional anti-de Sitter spacetime ($AdS_D$)
with $\bar{g}_{\mu\nu}(x)=\eta_{\mu\nu}$, as we will see in next section.

We now turn to the intermediate region, at which the scalar energy density
and the cosmological constant is comparable, between the core and the
asymptotic regions described by metrics (\ref{cores}) and (\ref{bbas}),
respectively. From the behavior of $g_{00}$, we may expect that a horizon is
located in the intermediate region. Since what we are looking for is the
existence of a black hole horizon, it will be convenient to write the metric
component $g_{00}$ in terms of $B(r)$ and $f(r)$ by eliminating $A(r)$ with
use of Eq. (\ref{ein3}) as
\begin{eqnarray}\label{e2ab}
e^{2A}B&=&C~r^{-\frac{2(d-2)}{(p+1)}}B^{-\frac{1}{(p+1)}}\nonumber\\
&&\times\exp\left[-\frac{2}{p+1}\int^r\left(\frac{8\pi G_Df^2-d+2}{r}
-\frac{2|\Lambda|-4\pi G_D(1-f^2)^2}{p+d-1}~r\right)\frac{dr}{B}\right],
\end{eqnarray}
where $C$ is an integral constant and it can be set to be $1$ with
the boundary condition $A=0$ at $r=0$. To examine the existence
and the property of the horizon, we begin assuming that there
exists a horizon at a finite coordinate distance $r_H$ from the
core or the asymptotic region, so that the timelike Killing vector
$\partial_t$ becomes null there, that is, $\exp[2A(r_H)]B(r_H)=0$.
Further we assume that $B(r)$ becomes zero and is analytic at the
horizon. However, we don't put further restriction on $A(r)$
because it could be singular at $r=r_H$, as can easily be observed
from Eq. (\ref{ein3}). We also assume that the scalar field $f(r)$
increases monotonically to unit value and behaves regularly at the
horizon. Then the position of the horizon $r_H$ and the value of
$f_H(\equiv f(r_H))$ are determined in a closed form from the
field equations.

Multiplying $e^{2A}B$ on Eq. (\ref{ein1}), eliminating $A'$ and $A''$ with
Eq. (\ref{ein3}), and taking the limit $r\to r_H$, we obtain
\begin{equation}\label{nhr1}
e^{2A(r_H)}\left[B_1+\Delta\right]\left[B_1+2\Delta\right]=0,
\end{equation}
where $B_1\equiv B'(r_H)$ and
\begin{equation}\label{Delta}
\Delta\equiv\frac{8\pi G_D f_H^2-d+2}{r_H}-
\frac{2|\Lambda|-4\pi G_D(1-f_H^2)^2}{p+d-1}r_H.
\end{equation}
On the other hand, eliminating $A'$, $A''$ from Eqs. (\ref{ein2}) and
(\ref{ein3}), and taking the limit $r\to r_H$, we get
\begin{equation}\label{nhr2}
\left[(p+2)B_1+2\Delta\right]\left[B_1+2\Delta\right]=0.
\end{equation}
There are two possibilities for $B_1$ satisfying simultaneously
the two relations Eqs. (\ref{nhr1}) and (\ref{nhr2}):
$B_1=-2\Delta\neq0$ or $B_1=0=\Delta$. However, the former
possibility should be excluded because with such condition the
metric component $g_{00}(r_H)$ does not vanish, but approaches to
a non-zero value at $r=r_H$, leading to a contradiction with
our starting assumption. This can easily be checked by inserting
following series expansions into Eq. (\ref{e2ab})
\begin{eqnarray}
B(r)&\approx&B_1(r-r_H)+B_2(r-r_H)^2+B_3(r-r_H)^3+\cdot\cdot\cdot,
\label{brh}\\
f(r)&\approx&f_H+f_1(r-r_H)+f_2(r-r_H)^2+\cdot\cdot\cdot. \label{frh}
\end{eqnarray}
The latter possibility only is acceptable and satisfies the
condition for the surface at $r=r_H$ to be a horizon. Inserting
above series expansions into Eq. (\ref{e2ab}) and using
$B_1=0=\Delta$ give up to the leading term
\begin{equation}\label{eab}
e^{2A}B\sim (r-r_H)^{\frac{1}{p+1}\left[-1+\frac{4}{B_H}\frac{2|\Lambda|-
4\pi G_D(1-f_H^2)^2}{p+d-1}\right]},
\end{equation}
where we have replaced $B_2$ with $B_H$, i.e. $B_2\equiv B_H$.
Therefore, if $B_H<4[2|\Lambda|-4\pi G_D(1-f_H^2)^2]/(p+d-1)$,
$e^{2A}B$ touches zero at $r=r_H$. We will see later that this condition is
always satisfied. Since $\Delta=0$, from Eq. (\ref{Delta}) we find the position
of the horizon in a closed form
\begin{equation}
r_H^2 = \frac{(p+d-1)(8\pi G_D f_H^2-d+2)}
         {2|\Lambda|-4\pi G_D(1-f_H^2)^2} \label{rh}.
\end{equation}
The fact $B'(r_H)=0$ indicates that the horizon is degenerated like that of
an extremal black brane. This tells that the black $p$-brane solution
with Ricci-flat world volume is extremal, as we have expected.

Inserting series expansions Eqs. (\ref{brh}) and (\ref{frh}) (with $B_1=0$)
into Eq. (\ref{ein3}) and using the condition $\Delta=0$, one finds that
\begin{equation}\label{napp}
A'(r)\approx
\frac{\alpha}{r-r_H}+O[(r-r_H)^0],
\end{equation}
where
\begin{eqnarray}
\alpha\equiv\frac{1}{p+1}\left[p+2-2\frac{2|\Lambda|-4\pi G_D(1-f_H^2)^2}
{(p+d-1)B_H} -p\frac{16\pi G_D(1-f_H^2)f_Hr_Hf_1}{(d-1)(p+d-1)B_H}\right],
\label{alph}
\end{eqnarray}
Here, the coefficients $B_H$, $f_H$, and $f_1$
can be determined in terms of parameters given in the theory using the
equations of motion.
Thus $A(r)$ is at most logarithmically divergent at $r=r_H$:
\begin{equation}\label{nr}
A(r)\approx -\alpha \ln(r-r_H) + O[(r-r_H)^0],
\end{equation}
and the $(00)$-component of the metric then becomes
\begin{equation}\label{g00}
g_{00}=e^{2A(r)}B(r)\approx B_H(r-r_H)^{2(1-\alpha)}.
\end{equation}
With series expressions of $B(r)$, $f(r)$ and $A(r)$ at the horizon,
from Eqs. (\ref{ein1}) and (\ref{ein3}) we obtain
\begin{equation}\label{bha}
B_H(1-\alpha)^2=\frac{2|\Lambda|-4\pi G_D(1-f_H^2)^2}{(p+1)(p+d-1)},
\end{equation}
and then simultaneously solving this and Eq. (\ref{alph}) for $B_H$ and
$\alpha$, we get
\begin{eqnarray}
1-\alpha&=&\frac{1}{2(2+\gamma)}\left[1\pm\sqrt{\frac{p+9+4\gamma}{p+1}}
             \right], \label{1-alpha}\\
 \nonumber\\
B_H &=&\frac{|\Lambda|-2\pi G_D(1-f_H^2)^2}{p+d-1}
\left[(p+5+2\gamma)\mp \sqrt{(p+1)(p+9+4\gamma)}~\right],\label{bh}
\end{eqnarray}
where
\begin{equation}
\gamma\equiv \frac{16\pi pG_Dr_H(1-f_H^2)f_Hf_1}
            {(d-1)(2|\Lambda|-4\pi G_D(1-f_H^2)^2)}.
\end{equation}
For lower sign, $(1-\alpha)$ is negative. This branch of solutions should be
excluded because, in that case, the norm of $\partial_t$ ($=e^{2N}B$) will
diverge at $r=r_H$, leading to a contradiction with our starting assumption
that $g_{00}$ is zero at $r=r_H$. We will discuss more about this branch
in next section. Whereas, with the upper sign $(1-\alpha)$ is positive saying
that the surface of $r=r_H$ is a horizon. $B_H$ satisfy our starting assumption
that $e^{2A}B\to 0$ as $r\to r_H$, as can be seen from Eq. (\ref{eab}).
The coefficient $f_1$ can be completely determined in terms of $r_H$ and $f_H$
using Eqs. (\ref{sfe}), (\ref{brh}), (\ref{frh}) and (\ref{napp}):
\begin{equation}
f_1=\frac{(d-1)f_H/r_H-(8\pi G_Df_H^2-d+2)}{r_H^2\left(p\displaystyle{
\frac{8\pi G_Df_H}{(p+d-1)r_H}}-f_H^2\right)}.
\end{equation}
On the other hand, considering limiting behaviors of the scalar field and the
metric functions at the horizon, from the scalar field equation Eq. (\ref{sfe})
we obtain another relation for $r_H$ and $f_H$
\begin{equation}\label{rh2}
r_H^2 =\frac{d-1}{1-f_H^2}.
\end{equation}
Solving simultaneously the two equations (\ref{rh}) and (\ref{rh2}), $r_H$ and
$f_H$ can be expressed in terms of parameters in the theory, $\Lambda$ and
$G_D$, as
\begin{eqnarray}
r_H^2&=&\frac{(p+d-1)(8\pi G_D-d+2)}{4|\Lambda|}
\left[1\pm\sqrt{1- \frac{32\pi G_D|\Lambda|(d-1)(2p+d-1)}
{(8\pi G_D-d+2)^2(p+d-1)^2}}~\right], \label{rh3}\\
f_H^2&=&\frac{p}{2p+d-1}+\frac{(p+d-1)}{8\pi G_D(2p+d-1)}\left[
\frac{}{}\right.(d-2)\nonumber\\
&&\left. \pm (8\pi G_D-d+2)\sqrt{1-\frac{32\pi G_D|\Lambda|(d-1)(2p+d-1)}
{(8\pi G_D-d+2)^2(p+d-1)^2}}~\right]\label{fh3}.
\end{eqnarray}
The complete determination of the position of the horizon $r_H$ and the value
of the scalar amplitude $f_H$ is an inevitable result since the leading term
of Einstein equations Eqs. (\ref{ein1})-(\ref{ein3}) and the scalar field
equation (\ref{sfe}) lead to two algebraic equations for $r_H$ and $f_H$.
The coefficients $B_3$ and $f_1$ are expressed in terms of $r_H$ and $f_H$ by
using the field equations, so $B_3$ and $f_1$ also are completely determined
in terms of the parameters of the theory. Since the higher order coefficients
of the series expansions (\ref{brh}) and (\ref{frh}) can be expressed in terms
of lower order coefficients, every coefficients can be determined completely
in closed forms. This implies that the flat brane solution with horizon is
unique for given parameters of the theory.

The size of the horizon ($r_H$) and the value of scalar field at the horizon
($f_H$) have two different expression for parameters given in the theory,
that is, we have two different types of black brane solutions.
For the first type of solutions with lower sign, the horizon is formed inside
the brane core and the bulk cosmological constant has to be super-Planckian
for generic values of symmetry breaking scale of the theory. The condition
that $f_H^2$ should be positive and real requires that the bulk cosmological
constant have a value within a limited region above the Planck scale:
\begin{equation}
2\pi G_D-\frac{(d-2)(p+d-1)}{2}<|\Lambda|<\frac{(8\pi G_D-d+2)^2(p+d-1)^2}
{32\pi G_D(d-1)(2p+d-1)}.
\end{equation}
For this value of the bulk cosmological constant, the core size of
a black brane will be larger than the size of the horizon. As an
example, for $p=3,~d=2$ the value of the scalar field at the
horizon is at most $\sqrt{3/7}$ showing that the core region
extends outward beyond the horizon. Then, solutions of this sort
may be viewed as small black branes lying within larger global
branes. However, these solutions may develop a classical
instability in a similar way to the case of the magnetically
charged Yang-Mills solitonic black holes with sufficiently small
horizon \cite{LNW}. This instability may lead to the possibility
that such a black brane could evaporate completely, leaving in its
place a nonsingular global defect. The question of whether the
black branes with small horizon lead to similar instabilities
discussed in Ref. \cite{LNW} itself will be an interesting one, but
is beyond the scope of this paper.

On the other hand, for solutions with upper sign the mass scale of
the bulk cosmological constant could be well below the Planck
scale and the horizon can be formed far outside the brane core for
generic values of the symmetry breaking scale and the bulk
cosmological constant, as can easily be checked in Eq. (\ref{fh3}).
Solutions of this sort will be matched smoothly to a version of
`thin-wall approximation' found in Ref. \cite{KMR}. In this paper,
we will consider only the black brane solutions corresponding to
the upper sign, because we are interested in only solutions with
horizon larger than brane core with respect to the brane world
scenario of Ref. \cite{KMR}.

The metric Eq. (\ref{bbm}) then has the form near $r=r_H$
\begin{equation}\label{bbhm}
ds^2\approx B_H[\sigma(r-r_H)]^{2(1-\alpha)}\bar{g}_{\mu\nu}(x)dx^\mu dx^\nu+
            \frac{dr^2}{B_H(r-r_H)^2}+r_H^2 d\Omega_{d-1}^2,
\end{equation}
where $\sigma$ is $+1$ for the exterior region ($r>r_H$) and $-1$ for the
interior region ($r<r_H$). Note that, so far, we have implicitly
assumed that we examine the near horizon region from the outside toward the
horizon at $r_H$ and hence obtained only exterior metric ($\sigma=1$).
Had we done from the outside region, we would then obtain the interior metric
($\sigma=-1$).

Up to now, finding an exact analytic solution to the field
equations Eqs. (\ref{ein1})-(\ref{sfe}) being impossible, we have
examined behavior of the solution of the global $p$-brane at a few
separated regions. The spacetime of the global black $p$-brane
will be described via metrics Eqs. (\ref{cores}), (\ref{bbas}) and
(\ref{bbhm}) provided that they are connected mutually. For metric
functions, $A(r)$ increases near the origin, diverges at $r_H$ and
decreases and rapidly approaches to a constant value at the asymptotic
region, as can be seen from Eqs. (\ref{arc}), (\ref{ara}) and (\ref{nr}).
$B(r)$ decreases near the origin, vanishes at $r_H$ and
increases quadratically at the asymptotic region, as indicated in
Eqs. (\ref{brc}), (\ref{bra}) and (\ref{brh}). The scalar field
$f(r)$ starts to grow linearly from zero at the origin, reaches to
a fixed value $f_H$ of Eq. (\ref{fh3}) and then approaches to its
vacuum value at the asymptotic region. From the behaviors of the
metric functions and scalar field, we can easily guess that the
fragments of solution can be connected mutually if they behaves
monotonically at each side of the horizon.

\begin{figure}[htbp]
\centerline{\epsfig{file=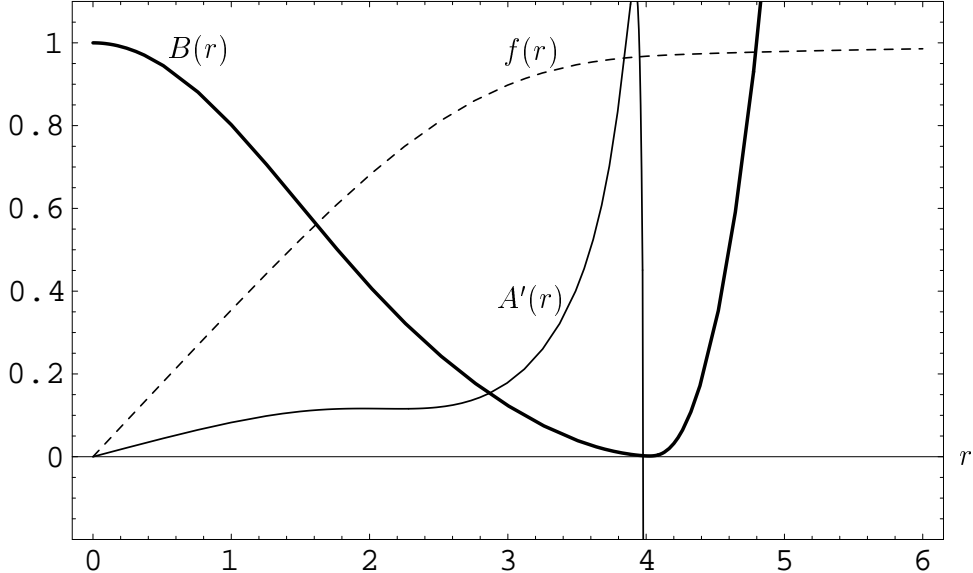, width=13cm}}
\caption{
A flat 3-brane solution surrounded by a degenerated horizon with two
transverse directions $(d=2)$ when $G_D=0.15$ and $|\Lambda| =0.420872997$.
With the thick solid line we report the behavior of the metric function $B(r)$
and with the dashed line the behavior of the scalar field $f(r)$, whereas with
the thin solid line the behavior of $A'(r)$.
}
\label{fig:abf}
\end{figure}

While we have not resolved yet whether the solutions obtained in
separated regions can be connected mutually or not, numerical
calculations indicate that the metric functions and the scalar
field behave monotonically at each side of the horizon, and the
solutions obtained in separated regions are connected smoothly
showing the existence of the black $p$-brane solution. Such an
example of the black brane solution is shown in FIG.1. The
obtained solution is for an extremal black 3-brane with two
transverse spatial direction $(d=2)$, that is, a global black
string-like defect. The scalar field $f(r)$ (the dashed line)
behaves monotonically for every $r$ and connects $f(0)=0$ and
$f(\infty)=1$ smoothly across the horizon at $r=r_H$. The metric
function $B(r)$ (the thick solid line) decreases monotonically
inside the horizon and vanishes at the horizon $r_H$, and then
increases monotonically. On the other hand, the derivative of the
metric function $A(r)$ (the thin solid line) shows a singular
behavior at the horizon as shown in Eq. (\ref{napp}).

Before ending this section, we briefly discuss the scales of
parameters of the theory: the symmetry breaking scale $v$, the Planck scale
$M_*$ and the cosmological constants $\Lambda$.
The right-hand side of Eq. (\ref{rh}) ought to be positive. This condition
yields $v^2>((d-2)/8\pi)~M_*^{D-2}$.\footnote{Since we are interested in
black brane solutions with horizon larger than the size of the brane core,
here we have assumed that $f_H\approx 1$. This assumption will be valid for
discussions on the spacetime structure of black brane solutions with horizon
outside the brane core.} While the condition does not give any new information
for $d=2$, for $d>2$, the condition sets the symmetry breaking scale $v$ to
be of the order of the fundamental scale $M_*$.
On the other hand, in order to trust the theory, $v^{2/D-2}$ ought to be
smaller than the fundamental scale: $v^{2/D-2}<M_*$. Hence, we assume that
$(d-2)/8\pi<v^2/M_*^{D-2}<1$. We also assume that $v^{2/D-2}<M_*$ even for
$d=2$, but this seems valid for naturalness reasons. The horizon size $r_H$
is then determined by the bulk cosmological constant $\Lambda$, viz,
$r_H^{-2}\sim |\Lambda|$.
There is no restriction on the scale of the cosmological
constant. It seems to be natural that the cosmological constant is of the
order of the fundamental scale in the absence of a mechanism to protect a
small cosmological constant from radiative corrections. However, given our
ignorance concerning the bulk cosmological constant problem, there is a priori
no reason to expect that $\Lambda$ is of the same order as the fundamental
scale $M_*^2$. Thus, we will treat the cosmological constant $\Lambda$ simply
as an input parameter.

\subsection{Non-extremal black branes: bent branes surrounded by two
horizons}\label{secSolsB}
~\\
In the previous subsection we found extremal black brane-like solutions.
Such extremal states are usually considered as critical limits of
non-extremal black objects. There may be various excitations upon the extremal
states. An example is the usual static non-extremal black $p$-branes found in
supergravity and superstring theories \cite{HS}. The non-extremality breaks
the bulk lorentz invariance \cite{kolb,poritz} in the brane world directions,
and the bulk curvature components in the brane world directions depend on the
extra dimension coordinates. Another example could be bent branes surrounded
by horizons among them, of which world volume has non-vanishing constant
curvature (i.e., $\hat{R}\neq0$). Such objects may be important, as it makes
the brane world scenario on the global defects more likely. We will try to
show the existence of such brane solutions. We find out series solutions at a
few separated regions as we did for flat branes in subsection \ref{secSolsA}.

At the origin, we find series solutions similar to those of the
Ricci-flat brane, Eqs. (\ref{frc}-\ref{brc}), up to the leading order.
The coefficients $A_2$ and $B_2$ are slightly modified with terms depending
on $\hat{R}$ as
\begin{eqnarray}
\bar{A}_2&\equiv&
\left[\frac{4\pi G_D}{d-1}\left(f_0^2-\frac{p}{2d(p+d-1)}\right)\right.
 \nonumber\\
&&\left. +\frac{p|\Lambda|}{d(d-1)(p+d-1)}+\frac{p(p+1)-d(d-1)}
{2(p+1)d(d-1)(p-d+1)}\hat{R}~\right],\label{a2b} \\
                                                   \nonumber\\
\bar{B}_2&\equiv&-\left[\frac{8\pi G_D}{d-1}\left(f_0^2
    -\frac{p-d+1}{2d(p+d-1)}\right) +\frac{2(p-d+1)|\Lambda|}{d(d-1)(p+d-1)}
           +\frac{\hat{R}}{d(d-1)}\right]\label{b2b}.
\end{eqnarray}
The metric at the core then takes the form of
\begin{equation}\label{coresb}
ds^2\approx\left[1-\left(\frac{4\pi G_D-2|\Lambda|}{d(p+d-1)}
-\frac{\hat{R}}{d(p+1)}\right) r^2\right]\hat{g}_{\mu\nu}(x)dx^\mu dx^\nu
+\frac{dr^2}{1+\bar{B}_2r^2}+r^2d\Omega_{d-1}^2,
\end{equation}
where $\hat{g}_{\mu\nu}(x)$ is a metric on the world volume of the brane,
which satisfies $(p+1)$-dimensional Einstein equation (\ref{ein4}) with
nonzero curvature $\hat{R}$. This shows that the metric component $g_{00}$
of the spacetime decrease in the transverse directions near the origin when
$4\pi G_D>2|\Lambda|+[(p+d-1)/(p+1)]\hat{R}$.

In the far region from the core, the metric functions behave in the same
manner as in flat-brane solutions up to the leading orders regardless the
curvature scale of the brane world volume, as can be easily seen from
Eq. (\ref{ein1}). The first term containing $\hat{R}$ of Eq. (\ref{ein1})
decays out and becomes negligible compared to other terms at the asymptotic
region, which approach to a constant values. The metric is then simply
\begin{equation}\label{nbbas}
ds^2\approx B_{\infty}r^2~\hat{g}_{\mu\nu}(x)dx^\mu dx^\nu
+\frac{dr^2}{B_{\infty}r^2}+r^2d\Omega_{d-1}^2,
\end{equation}
where $B_\infty$ is the same coefficient with that of the metric
(\ref{bbas}).

As in the case of flat brane (if they are formed) the horizons
will be formed in the intermediate region between the core and the
asymptotic regions. We will probe the horizons from both the core
and the asymptotic regions. We assume that there exist horizons at
finite coordinate distances $r=r_H$ from observers living on the core
or in the asymptotic region. We will follow the same
procedure as done in the flat brane case to identify the location
of and the properties of the horizon. While Eq. (\ref{nhr2}) is not
modified with inclusion of non-zero curvature $\hat{R}\neq0$, the
relation (\ref{nhr1}) is modified as
\begin{equation}\label{nhr}
2\hat{R}=e^{2A(r_H)}\left[B_1+\Delta\right]\left[B_1+2\Delta\right].
\end{equation}
The only possibility satisfying simultaneously two equations
(\ref{nhr2}) and (\ref{nhr}) is $B_1=-2\Delta/(p+2)\neq0$.
Inserting this into Eq. (\ref{nhr}) gives
\begin{equation}\label{e2ab2}
e^{2A(r_H)}B_1^2=\frac{4\hat{R}}{p(p+1)}.
\end{equation}
This implies that $B_1\neq0$ unless $\hat{R}=0$, and the horizons are not
degenerated contrary to the case of flat branes.
Since the left hand side is positive definite, the world volume curvature
has to be positive constant, that is, the corresponding $(p+1)$-dimensional
brane world has to be de Sitter spacetime. On the other hand, inserting series
expansions Eqs. (\ref{brh}) and (\ref{frh}) into Eq. (\ref{e2ab}) and taking the
limit $r\to r_H$ we obtain
\begin{equation}\label{e2A}
e^{2A(r_H)}=r_H^{-\frac{2(d-2)}{p+1}}|B_1|^{-\frac{p+2}{p+1}},
\end{equation}
where we have set $C=1$ in Eq. (\ref{e2ab}) with the boundary condition
$A=0$ at $r=0$. From Eqs. (\ref{e2ab2}) and (\ref{e2A}) we get the absolute
value of $B_1$ at $r=r_H$
\begin{equation}\label{b1}
|B_1|=r_H^{\frac{2(d-2)}{p}}\left[\frac{4\hat{R}}{p(p+1)}\right]^{(p+1)/p}.
\end{equation}
Then, using the relation $B_1=-2\Delta/(p+2)$ we obtain following two relations
between $r_H$ and $f_H$:
\begin{equation}\label{rhfh}
\frac{2|\Lambda|-4\pi G_D(1-f_H^2)^2}{p+d-1}r_H^2-(8\pi G_Df_H^2-d+2)=
\pm\frac{p+2}{2}~r_H^{\frac{2(d-2)}{p}+1}\left[\frac{4\hat{R}}{p(p+1)}
\right]^{(p+1)/p},
\end{equation}
where the positive sign of the right hand side corresponds to $B_1>0$ and the
negative sign to $B_1<0$.
We need another relation between $r_H$ and $f_H$ to identify the position of
the horizon and the value of the scalar field at the horizon. However, unlike
the case of the flat brane the set of equations of motion doesn't give the
additional relation. The leading term of the scalar field equation (\ref{sfe})
that gave the additional relation in the case of the flat brane does role only
to determine $f_1$ in terms of $r_H$ and $f_H$.\footnote{$f_1$ is determined
in terms of $r_H$ and $f_H$ as $f_1=2f_H[d-1-(1-f_H^2)^2r_H^2]/
(p+2)r_H^2|B_1|$.}
So we cannot determine completely $r_H$ and $f_H$ unlikely in the case of the
flat brane.

Since for each sign the relation Eq. (\ref{rhfh}) admits a positive real root
of $r_H$ for a given $f_H$, we have two distinct values of $r_H$, i.e.,
${r_H}_+$ (for $+$ sign of r.h.s.) and ${r_H}_-$ (for $-$ sign of
r.h.s.).\footnote{Here, we assume that $|\Lambda|>2\pi G_D(1-f_H^2)^2$.
Note that this condition was automatically
satisfied for flat brane solutions as guaranteed by Eq. (\ref{bha}).
If we consider the flat brane solution as a critical case of a
bent brane solution, then the assumption will be natural. Remind
also that $8\pi G_Df_H^2>d-2$ from Eq. (\ref{rh}).} Therefore, we
have two distinct solutions near the surfaces at ${r_H}_\pm$
\begin{equation}\label{bnhm}
ds^2_\pm\approx e^{2{A_H}_{\pm}}{B_1}_{\pm}(r-{r_H}_\pm)~d\hat{s}^2
+\frac{dr^2}{{B_1}_{\pm}(r-{r_H}_\pm)}+{r_H}_\pm^2d\Omega_{d-1}^2,
\end{equation}
where $d\hat{s}^2\equiv \hat{g}_{\mu\nu}dx^\mu dx^\nu$, and
${A_H}_{\pm}$ and ${B_1}_{\pm}$ are values at ${r_H}_+$ and
${r_H}_-$, respectively. In usual, this type of metrics develops
curvature singularities at $r={r_H}_\pm$ except the case $p=0$.
However, as we will see in the next subsection, the spacetime
described by this metric is regular at $r={r_H}_\pm$ thanks to the
relation (\ref{e2ab2}) and so the surfaces at $r={r_H}_\pm$ are
coordinate singularities. Moreover, the fact that ${B_1}_+>0$ but
${B_1}_-<0$ seems to imply that surfaces at ${r_H}_+$ and
${r_H}_-$ correspond to the outer horizon and the inner horizon of
a non-extremal black brane, respectively. This will be true, if
${r_H}_+>{r_H}_-$ and the two solutions are mutually connected.
The reason is as follows. Since $e^{2A}$ is finite at the horizons
and is positive always, the spacetime structure is determined
solely by the behavior of $B(r)$. Since $B(r)$ is positive near
the core and at the asymptotic region, $B(r)$ has to be negative
at an intermediate region if there should be two horizons. Then
the tangent of $B(r)$ should be negative at the inner horizon and
positive at the outer horizon, as we have observed in the case of
the Reissner-Nordstr\"{o}m black hole.

If the horizons are far outside the core, then the scalar field has nearly the
same values at the inner and outer horizons (i.e., ${f_H}_-\simeq{f_H}_+
\simeq1$), and then we can easily see that ${r_H}_+>{r_H}_-$ from
Eq.(\ref{rhfh}). In particular, when $d=2$, this can be explicitly observed,
because the roots can be written in closed form as follows
\begin{eqnarray}
{r_H}_{\pm}&=&\pm\frac{(p+1)(p+2)}{8[|\Lambda|-2\pi G_D(1-{f_H}_\pm^2)^2]}
\left[\frac{4\hat{R}}{p(p+1)}\right]^{\frac{p+1}{p}} \nonumber\\
& &+\sqrt{\frac{(p+1)^2(p+2)^2}{64[|\Lambda|-2\pi G_D(1-{f_H}_\pm^2)^2]^2}
\left[\frac{4\hat{R}}{p(p+1)}\right]^{2\frac{p+1}{p}}+\frac{(p+1)
8\pi G_D{f_H}_\pm^2}{2[|\Lambda|-2\pi G_D(1-{f_H}_\pm^2)^2]}}.
\end{eqnarray}
Clearly, if ${f_H}_-\simeq{f_H}_+\simeq f_H$, then ${r_H}_+>{r_H}_-$.
The difference between ${r_H}_+$ and ${r_H}_-$ is given by
\begin{equation}
{r_H}_+-{r_H}_-\simeq\frac{(p+1)(p+2)}{4[|\Lambda|-2\pi G_D(1-{f_H}^2)^2]}
\left[\frac{4\hat{R}}{p(p+1)}\right]^{\frac{p+1}{p}}>0.
\end{equation}
The two horizons are degenerated in the limit $\hat{R}\to 0$, and
the position of the degenerated horizon coincides with that of the
extremal black brane given by Eq. (\ref{rh}). Therefore, our
solution seems to behaves correctly in the critical limit that
$\hat{R}\to0$, as expected from the usual non-extremal black
objects.

\begin{figure}[htbp]
\centerline{\epsfig{file=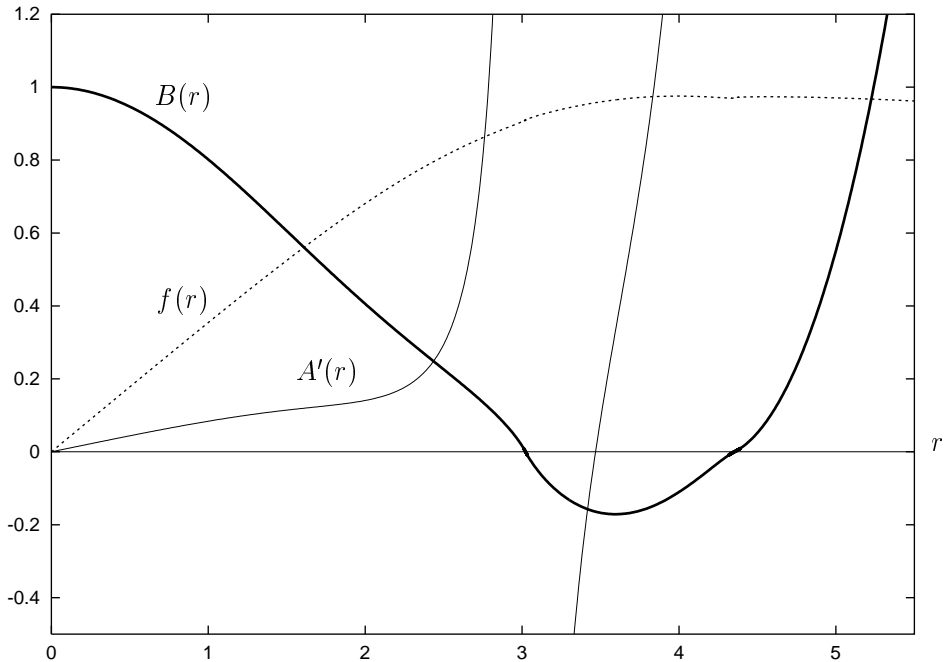, width=13cm}}
\caption{
A non-extremal black 3-brane solution with two transverse directions $(d=2)$
when $G_D=0.15$, $|\Lambda| =0.420872997$ and $\hat{R}=3\times10^{-8}$.
With the thick solid line we report the behavior of the metric function $B(r)$
and with the dashed line the behavior of the scalar field $f(r)$, whereas with
the thin solid line the behavior of $A'(r)$.
}
\label{bbb}
\end{figure}

The remaining step is now to see whether the solutions obtained in
separated regions can be mutually connected or not. The numerical
works indicate that the possibility of the existence of a smooth
configuration to connect the solutions (\ref{coresb}),
(\ref{nbbas}), and (\ref{bnhm}), and a smooth scalar field
configuration to interpolate between $f(0)=0$ and $f(\infty)=1$.
An example of such bent brane solution surrounded by two horizons
is illustrated in Fig. \ref{bbb}.

\section{Spacetime structures}\label{secSt}

\subsection{Extreme global black branes}\label{secStA}
~\\
In this subsection we discuss the spacetime structure induced via mutually
connected metrics (\ref{cores}), (\ref{bbas}) and (\ref{bbhm}).
We begin by examining the geometry inside the near horizon region.
As mentioned in the previous section, outside the core the behavior of metric
functions is controlled by the relative magnitude of the the scalar field
energy density $8\pi G_D/r^2$ to the cosmological constant $|\Lambda|$.
The horizon occurs in the region where $8\pi G_D/r^2$ and $|\Lambda|$ are
comparable. The spacetime of such region is well described by the metric
(\ref{bbas}).
The shape of the interior region will depend on value of the
cosmological constant. When the cosmological constant is of the same order
as the fundamental scale, both the core radius
$r_c\sim (\sqrt{\lambda}v)^{-1/2}$ and the horizon size $r_H$ are of the
order of the fundamental size $M_*^{-1}$ and so the scalar field energy
density $8\pi G_D/r^2$ and the cosmological constant $|\Lambda|$ are of
comparable magnitude over the entire interior region. Thus the interior
region is well described by the two metrics Eqs. (\ref{cores}) and (\ref{bbas}).

However, when the cosmological constant is much less than the
fundamental scale, the geometry of the interior region is more
complicated. In this case, the size of the horizon is much larger
than the size of the core. The core region and the near horizon
region are squeezed into small portions and, in between the two,
most portion is occupied by an intermediate region characterized
by $8\pi G_D/r^2\gg |\Lambda|$. Even though we don't have any
analytic solution for the intermediate region between the core and
the horizon, we could expect its geometry to be approximated by
that of the global defect spacetime in the absence of the negative
cosmological constant. The spacetime geometry depends on the
transverse space dimensions. When $d=2$, the geometry of this
region will resemble that of the Cohen-Kaplan solution \cite{CK2} with
removed curvature singularity. If $d\geq3$, the geometry of this region
would become similar to that of the global monopole solution of
Ref. \cite{Vil}.

The exterior region interpolates between the near horizon region
and the asymptotic region described by metrics (\ref{bbas}) and
(\ref{bbhm}), respectively. Since in the far region from the core
the cosmological constant dominates over the scalar field energy
density, one may expect the spacetime asymptotes to a negative
constant curvature space, which is $D$-dimensional anti-de Sitter
space $(AdS_D)$ with $\bar{g}_{\mu\nu}=\eta_{\mu\nu}$. The Riemann
curvature tensor corresponding to the metric (\ref{bbas}) is
\begin{equation}
R_{MNPQ}=\left(-B_{\infty}+\left[\frac{1}{r^2}\right]_{
M,N,P,Q\in \{\theta_a\}}\right)\left(g_{MP}g_{NQ}-g_{MQ}g_{NP}\right),
\end{equation}
where the second term in the bracket is non-vanishing only for components
$R_{\theta_a,\theta_b,\theta_c,\theta_d}$. Thus we easily see that, when
$d=2$, the spacetime described by the metric (\ref{bbas}) is a negative
constant curvature space, {\it i.e.}, an anti-de Sitter space with
curvature scale $\sqrt{B_{\infty}}$. While, for $d\geq3$ the space
is no longer a constant curvature space, but asymptotes to the
$D$-dimensional anti-de Sitter space $(AdS_D)$, as $r\to\infty$.
This type of spacetimes has been found in Refs. \cite{Gre,Vil}. In fact,
the metric (\ref{bbas}) can be rewritten to the form found in
\cite{Gre,Vil} introducing the proper radial distance $\xi\equiv
\int^rdr'/\sqrt{B(r')}$ as
\begin{equation}\label{amet}
ds^2\approx e^{2\sqrt{B_\infty}\xi}\bar{g}_{\mu\nu}(x)dx^\mu dx^\nu+d\xi^2
+e^{2\sqrt{B_\infty}\xi}d\Omega_{d-1}^2.
\end{equation}

\begin{figure}[htbp]
\centerline{\epsfig{file=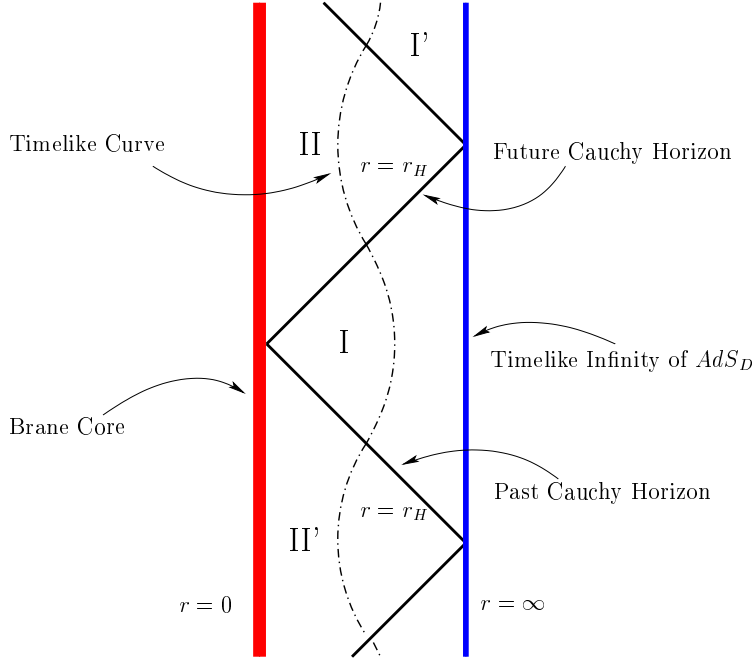,width=10cm}} \caption{
Penrose diagram for the extreme global black $p$-brane with field
configurations shown in FIG. \ref{fig:abf}. The spatial coordinates
$x^i$ and $\theta^a$ are suppressed. This diagram is very similar
to that of the extreme Reissner-Nordstr\"{o}m black hole, but now
there is a non-singular brane core at $r=0$ rather than a
curvature singularity. The boundary at $r\to\infty$ is timelike
rather than null. } \label{penrose}
\end{figure}

Finally, we examine the near horizon geometry.
We begin by studying the motion of test particles.
Due to the Lorentz invariance in the longitudinal direction with
$\bar{g}_{\mu\nu}(x)=\eta_{\mu\nu}$ and the spherical symmetry in the
transverse direction, it is sufficient to study null and timelike geodesics
in the system
\begin{eqnarray}
ds^2=-B_H[\sigma(r-r_H)]^{2(1-\alpha)}dt^2+\frac{dr^2}{B_H(r-r_H)^2}.
\end{eqnarray}
Defining the covariant 4-velocity
$u_{\hat{\mu}}=g_{\hat{\mu},\hat{\nu}}dx^{\hat{\nu}}/d\tau$,
we find the geodesic equation
\begin{eqnarray}
\kappa=g_{\hat{\mu},\hat{\nu}}u^{\hat{\mu}}u^{\hat{\nu}}
=-B_H[\sigma(r-r_H)]^{2(1-\alpha)}\left(\frac{dt}{d\tau}
\right)^2+\frac{1}{B_H(r-r_H)^2}\left(\frac{dr}{d\tau}\right)^2,
\end{eqnarray}
where $\tau$ is an Affine parameter (for timelike geodesics, it is
the proper time). Here $\kappa$ is 0 for null geodesics and $-1$
for timelike geodesics. There is a constant of the motion
\begin{eqnarray}
\epsilon =- g_{\hat{\mu},\hat{\nu}}\xi^{\hat{\mu}}u^{\hat{\nu}}
=-g_{00}\frac{dt}{d\tau}=B_H[\sigma(r-r_H)]^{2(1-\alpha)}\frac{dt}{d\tau},
\end{eqnarray}
where $\xi^{\hat{\nu}}\equiv (\partial/\partial t)^{\hat{\nu}}$ denotes the
static Killing vector. The equation motion then is
\begin{eqnarray}
\left(\frac{dr}{d\tau}\right)^2-\kappa B_H[\sigma(r-r_H)]^2
-\epsilon^2[\sigma(r-r_H)]^{2\alpha}=0.
\end{eqnarray}
The equation of motion tells that $(dr/d\tau)\to 0$ as $r\to r_H$.
The proper distance from a point $r_0$ to the horizon on a constant time slice
is logarithmically divergent as
\begin{eqnarray}
\chi(r)\sim \frac{1}{\sqrt{B_H}}\ln\left|\frac{r-r_H}{r_0-r_H}\right|.
\end{eqnarray}
However, test particles starting at a point $r_0$ and moving toward the
horizon reach $r=r_H$ and $t=\infty$ in the finite Affine parameter
\begin{eqnarray}
\tau=\int^{r_H}_{r_0}\frac{dr'}{\left\{\epsilon^2[\sigma(r'-r_H)]^{2\alpha}
+\kappa B_H(r'-r_H)^2\right\}^{1/2}}<\infty,
\end{eqnarray}
and, for null geodesics ($\kappa =0$), we explicitly get
\begin{equation}\label{npt}
\tau 
\buildrel r\to r_H\over\longrightarrow
\frac{1}{\epsilon}~\frac{[\sigma(r_0-r_H)]^{1-\alpha}}{1-\alpha}.
\end{equation}
Therefore, the coordinates $\{x_\mu,r,\theta_a\}$ are not geodesically
complete in both interior and exterior regions. Hence, we need
to extend each region onto new patches across the horizon. It is
well known that as the extension gets across a Cauchy horizon, it is
not unique. That is, depending on the choice of spacetime
identification there are different possibilities to analytically
continue the spacetime across the horizon. For an example, we
would consider extensions in which the interior region is
reflected across a Cauchy horizon so as to have its copy in the
next patch. Since the extension has its own horizon, such
extension can be repeated giving rise to an infinite array of the
interior region. In the same manner, we may obtain an infinite
array of the exterior region, which is a maximal analytic
extension of the exterior region. Another plausible extension is
the covering space of the black $p$-brane system, which is an
infinite lattice of the exterior regions and the interior regions
as depicted in FIG. \ref{penrose}. This spacetime is obtained by
repeating infinitely the extension procedure of the exterior
region to the interior region across the Cauchy horizon and its
inverse. This possibility involves no identifications
and thus contains no closed timelike curves (CTCs). This type of
extension seems to be more natural if we consider the field
configuration found in FIG. \ref{fig:abf}, where the scalar field
monotonically interpolates across horizon between the false vacuum
at the core and the true vacuum at asymptotic region. This
spacetime has a causal structure similar to that of the extreme
Reissner-Nordstr\"{o}m (RN) black hole in four-dimensional
spacetime, besides that it is perfectly regular everywhere, while
that of the extreme RN black hole has a timelike singularity at
the origin. The spacetime of the global black brane asymptotes to
a $D$-dimensional anti-de Sitter space, while that of the extreme
RN black hole to a flat Minkowski spacetime.

The near horizon geometry is very similar to that of the extreme
RN black hole, of which geometry has the topology of $AdS_2\times
S^2$ with constant curvature scales of $1/r_H$. It is easy to see
that the metric (\ref{bbhm}), with
$\bar{g}_{\mu\nu}(x)=\eta_{\mu\nu}$, describes a constant
curvature spacetime with topology of $AdS_{p+2}\times S^{d-1}$
with curvature scales $\sqrt{(1-\alpha)^2B_H}$ and $1/r_H$ for
$AdS_{p+2}$ and $S^{d-1}$, respectively: Components of the Riemann
curvature tensor corresponding to the metric (\ref{bbhm}) are
given by
\begin{eqnarray}
R_{\hat{\mu}\hat{\nu}\hat{\lambda}\hat{\rho}}&=&-(1-\alpha)^2B_H(g_{\hat{\mu}
\hat{\lambda}}g_{\hat{\nu}\hat{\rho}}-g_{\hat{\mu}\hat{\rho}}
g_{\hat{\nu}\hat{\lambda}}),  \\
R_{\theta_a\theta_b\theta_c\theta_d}&=&\frac{1}{r_H^2}(g_{\theta_a\theta_c}
g_{\theta_b\theta_d}-g_{\theta_a\theta_d}g_{\theta_b\theta_c}),
\end{eqnarray}
where $\hat{\mu},\hat{\nu},...$ refer to $x_\mu$ and $r$, and
$\theta_a,\theta_b,...$ do indices on $S^{d-1}$.

It will be helpful to rewrite the metric (\ref{bbhm}) in the form of
Robinson-Bertotti (RB) metric as
\begin{equation}\label{rbm}
ds^2_{RB}\approx \frac{\rho^2}{\rho_H^2}\eta_{\mu\nu}dx^\mu dx^\nu
           +\frac{\rho_H^2}{\rho^2}d\rho^2+r_H^2d\Omega_{d-1}^2,
\end{equation}
where we have introduced the Poincar\'{e} radial coordinate
\begin{eqnarray}\label{rho}
\frac{\rho}{\rho_H}=\sqrt{B_H}[\sigma(r-r_H)]^{1-\alpha}~~\Leftrightarrow~~
r=r_H+\sigma\left(\frac{\rho}{\sqrt{B_H}~\rho_H}\right)^{1/(1-\alpha)},
\end{eqnarray}
with $\rho_H\equiv 1/\sqrt{B_H}(1-\alpha)$.
If the radial coordinate $\rho$ ranges between $0$ and $\infty$, then the
Poincar\'{e} coordinates $\{x_\mu,\rho\}$ cover only one half of $AdS_{p+2}$,
which is named as the Poincar\'{e} patch. In this coordinate system, the
Cauchy horizon is at $\rho=0$ and the boundary of $AdS_{p+2}$ at $\rho=\infty$
corresponds to a $(p+1)$-dimensional flat Minkowski spacetime, $R^{1,p}$.
However, with the definition Eq. (\ref{rho}) $\rho$ has restricted ranges
and so each region covers only a part of the Poincar\'{e} patch.
When $(1-\alpha)$ is positive, $\rho$ ranges from $0$ to a finite value
$\rho_f$. For the interior solution ($\sigma=-1$), roughly $\rho_f\sim
r_H^{1-\alpha}/(1-\alpha)$ (at $r\sim 0$). For the exterior region
($\sigma=1$), the metric (\ref{rbm}) should be replaced
with asymptotic solution (\ref{bbas}) or (\ref{amet}) with appropriate
coordinate transformation at $\rho_f$. Hence, with positive $(1-\alpha)$,
the metric (\ref{rbm}) describes portions including the AdS horizon of
$AdS_{p+2}\times S^{d-1}$.

On the other hand, when $(1-\alpha)$ is negative, the boundary at
$r=r_H$ is not a horizon and the metric is not that of the near
horizon geometry of extremal black brane because the timelike
Killing vector $\partial_t$ diverges at $r=r_H$. And, as can be
observed from Eq. (\ref{npt}), the boundary is at a affinely
infinite distance. Thus, the boundary at $r=r_H$ corresponds to
the timelike infinity of $AdS_{p+2}$, which is conformal to flat
Minkowski spacetime ${\cal M}^{1,p}$. In this branch, $\rho$ ranges from
a finite value to infinity. Hence, the metric (\ref{rbm})
describes an AdS fragment including the $(p+1)$-dimensional flat
Minkowski boundary but cutting out the AdS horizon.

Recall that near the horizon of the extremal RN black hole the asymptotic
metric is also written in the form of the RB metric as
\begin{equation}
ds_{RB}^2= -\frac{\rho^2}{r_H^2}dt^2+\frac{r_H^2}{\rho^2}d\rho^2
+r_H^2d\Omega_2^2,
\end{equation}
where $r_H=GM$ is the size of the horizon of the extremal RN black hole.
It is a kind of \lq Kaluza-Klein' vacuum in which two directions are
compactified and the \lq effective' spacetime is the two-dimensional $AdS_2$
spacetime of constant negative curvature.
Similarly, the $AdS_{p+2}\times S^{d-1}$ spacetime described by the near
horizon metric (\ref{bbhm}) or (\ref{rbm}) is also a \lq Kaluza-Klein' vacuum.
Usually, such type of vacuum is realized as a supersymmetric bosonic
configuration, {\it i.e.}, a \lq supersymmetric' vacuum of a
higher-dimensional supergravity theory. It also appears as the near horizon
geometry of non-dilatonic black $p$-brane solutions of higher-dimensional
supergravity or superstring theories. It will be interesting to compare
the metric (\ref{bbhm}) with that of supersymmetric non-dilatonic black
$p$-branes.

For comparison, we write down the metric of non-dilatonic extremal black
$p$-brane:
\begin{equation}\label{pmetric}
ds^2={[\sigma\Delta(r)]}^{\frac{2}{p+1}}\eta_{\mu\nu}dx^\mu dx^\nu
     +\frac{dr^2}{{\Delta(r)}^2}+r^2 d\Omega_{d-1}^2,
\end{equation}
where the extra dimension is fixed to be $d=4p/(p-1)$ and
\begin{equation}
\Delta(r)\equiv 1-\left(\frac{r_H}{r}\right)^{d-2}.
\end{equation}
In the limit $r\to r_H$, the metric is written as
\begin{eqnarray}\label{npmetric}
ds^2&\approx&\left[\frac{2(p+1)}{(p-1)r_H}\right]^{\frac{2}{p+1}}
 [\sigma(r-r_H)]^{\frac{2}{p+1}}\eta_{\mu\nu} dx^\mu dx^\nu
+\left[\frac{2(p+1)}{(p-1)r_H}\right]^{-2}\frac{dr^2}{(r-r_H)^2}
 +r_H^2d\Omega_{d-1}^2,
\end{eqnarray}
which explicitly has the same form as Eq. (\ref{bbhm}).
Even though Eqs. (\ref{bbhm}) and (\ref{npmetric}) describe the same
geometry near the horizon, the global spacetime structures of the two system,
the global black brane and the non-dilatonic black brane, are much different.
The spacetime of the global black brane is completely regular everywhere,
whereas that of the non-dilatonic black brane is singular at the center $r=0$.
The geometry of the spacetime of the global black brane asymptotes to an $AdS$
space, while that of the non-dilatonic brane does to the Minkowski spacetime.

Notice that if we used a metric ans\"{a}tz using a proper radial
distance instead the Schwarzschild coordinates, we would obtain a
spacetime described Eq. (\ref{bbhm}) as an asymptotic solution because
the horizon is sitting at infinite proper distance from the core.
This is the thing done in Refs. \cite{Gre} and \cite{Vil}, in which the
authors discovered cigar-like warped spacetimes which is nothing but the
infinitely long AdS throat with topology of $AdS_{p+2}\times
S^{d-1}$. It is easy to see relation between the near horizon
solution and the cigar-like warped spacetime obtained in
Refs. \cite{Gre,Vil}: Introducing a new radial coordinate
$\chi$($>0$) such that
\begin{equation}
\exp(-k\chi)\equiv \sqrt{B_H}~[\sigma(r-r_H)]^{1-\alpha},
\end{equation}
the metric Eq.(\ref{bbhm}) is rewritten as
\begin{equation}\label{grg}
ds^2\approx \exp\left(-2k\chi\right)\bar{g}_{\mu\nu}(x)dx^\mu dx^\nu
            +d\chi^2+r_H^2 d\Omega_{d-1}^2,
\end{equation}
where the curvature scale $k$ of the $AdS_{p+2}$ is defined by
$k\equiv(1-\alpha)\sqrt{B_H}$. For the interior solution ($\sigma
=-1$), $\chi$ runs from a finite value to $\infty$ (at $r=r_H$).
For the exterior solution ($\sigma =1$), $\chi$ runs between
$\infty$ (at $r=r_H$) and $-\infty$ (at $r=\infty$), but $\chi$
ought to be truncated at a finite distance as, sufficiently large
$r$, the near horizon geometry is replaced by the asymptotic
spacetime Eq. (\ref{amet}). The metric (\ref{grg}) coincides with
the cigar-like warped solution of Refs. \cite{Gre,Vil} in the thin
core limit of $f_H\to 1$ where the horizon locates outside the
core. Hence, among solutions obtained in \cite{Gre} and
\cite{Vil}, the meaningful solutions relevant for the
Randall-Sundrum scenario would quite naturally be
interpreted as the near horizon geometry of extremal global black
$p$-branes. Moreover, we also find that the two solutions,
Eqs. (\ref{amet}) and (\ref{grg}), which were apparently treated
disjointedly in Refs. \cite{Gre,Vil}, can be matched to each other
and describe the exterior region together.

\begin{figure}[htbp]
\centerline{\epsfig{file=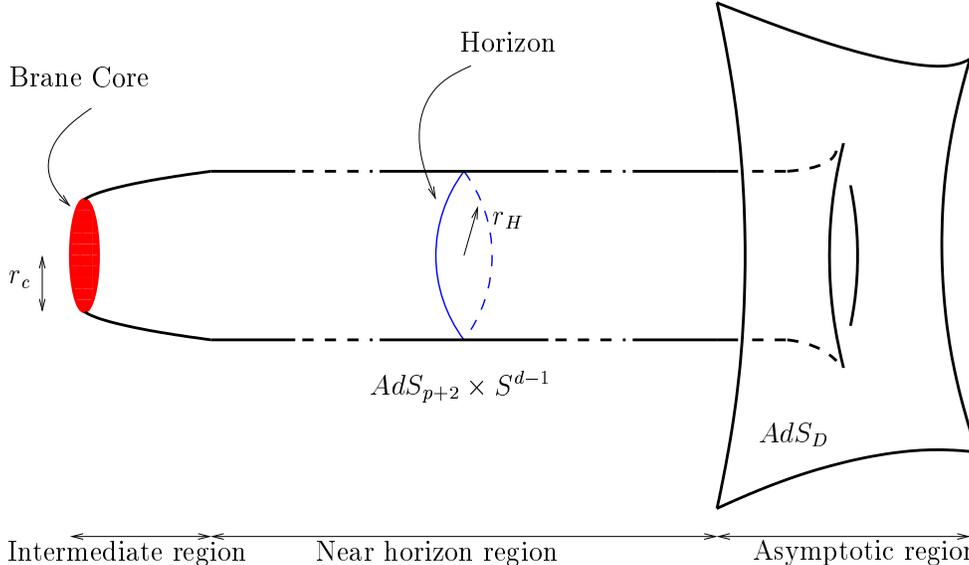, width=13cm}}
\caption{
The shape of the slice of the transverse part of the spacetime of the extreme
global black brane. The time and longitudinal directions are suppressed.
Circles parallel to the horizon line represent $S^{d-1}$.
The surface is a transversal slice.
}
\label{condom}
\end{figure}

In summary, the exterior region outside the horizon interpolates
between the near horizon region, the infinitely long AdS throat
with topology of $AdS_{p+2}\times S^{d-1}$, and the asymptotic
region which asymptotes to $AdS_d$. The interior region can be
divided to three separated regions, core region, near horizon
region and intermediate region between the two regions.
The near horizon region corresponds to the infinitely
long AdS throat $(AdS_{p+2}\times S^{d-1})$. As such, the central
region surrounded by the AdS throat region looks like a one-sided
Randall-Sundrum domain-wall embedded in $(p+2)$-dimensional AdS
spacetime $(AdS_{p+2})$. When the cosmological constant is of the
same order as the fundamental scale, i.e.,
$M_*\sim|\Lambda|^{1/2}\sim M_{pl}$, most portion of the interior
region is occupied by the core and near horizon regions and so the
interior region is well described by only the two region. The
curvature radii of both $AdS_{p+2}$ and $S^{d-1}$ are of the order
of the Planck scale. At low-energy below the Planck scale,
the extra space is reduced effectively to a one-dimensional space.
Consequently, the global $p$-brane core looks like a
$p$-dimensional domain-wall embedded in an $AdS_{p+2}$ bulk
spacetime. In the \lq thin core-approximation' limit, the system
is essentially the same as that of the original Randall-Sundrum
scenario. However, when the cosmological constant is much less
than the fundamental scale, the core and the near horizon regions
are squeezed into small portions and, in between the two most
portion is occupied by an intermediate region. Since the
cosmological constant is negligible in the intermediate region,
its geometry can be approximated by that of the global defect
spacetime in the absence of the negative cosmological constant.
The geometry of this region resembles that of the Cohen-Kaplan
solution \cite{CK2} as $d=2$ and the global monopole solution
\cite{Vil} as $d\geq3$. Putting the above results together, the
spacetime geometry of the global black $p$-brane is illustrated in
FIG. \ref{condom}.

\subsection{Non-extreme global black branes}\label{secStB}
~\\
In this subsection we discuss the spacetime of the bent brane surrounded by
two horizons, of which a field configuration is shown in FIG. \ref{bbb}.
Since the world volume curvature of the brane core has to be positive constant,
the corresponding world volume metrics are those of $(p+1)$-dimensional de
Sitter spacetime, e.g.,
\begin{eqnarray}
d\hat{s}^2 
&=&-dt^2+e^{2Ht}dx_i^2 \label{fum}\\
&=&-(1-H^2\gamma^2)dt^2+(1-H^2\gamma^2)^{-1}d\gamma^2
   +\gamma^2d\Omega_{p-1}^2, \label{bhm}
\end{eqnarray}
where $\gamma^2\equiv x_i^2$, and $H$ is the expansion rate along
the brane. The first metric (\ref{fum}) is defined in terms of
planar coordinates and corresponds to a flat Friedman universe.
The second one is static and is the analog of the Schwarzschild
metric for the flat de Sitter spacetime. The $t$ in these
coordinates is not the same as the $t$ in planar coordinates, but
we are running out of letters. In this coordinate system
$\partial_t$ is a Killing vector and generates the time
translation symmetry. From (\ref{bhm}), we see that at
$\gamma=H^{-1}$ the norm of $\partial_t$ vanishes, so that it
becomes null. Thus, the surface at $\gamma=H^{-1}$ is the usual de
Sitter horizon. On the other hand, since $H$ is related to the
world volume curvature as $\hat{R}=p(p+1)H^2$, from
Eqs. (\ref{e2ab2}) and (\ref{e2A}) it can be read off that
\begin{equation}\label{HH}
H=\pm \frac12 e^{{A_H}_\pm}|{B_1}_\pm| .
\end{equation}

The metrics (\ref{coresb}), (\ref{nbbas}), and (\ref{bnhm})
describe only a few separated regions, but they are smoothly
connected as shown in FIG. \ref{bbb}. Similarly to the extreme
black brane case, when the cosmological constant is of the order
of the fundamental scale, both the core size $r_c$ and the sizes
of the horizons ${r_H}_\pm$ are of the order of $M_*^{-1}$ and so
the spacetime of the interior region inside ${r_H}_-$ will be well
described by the two metrics (\ref{coresb}) and (\ref{bnhm}). On
the other hand, when the cosmological constant is much less than
the fundamental scale, in the intermediate region the transverse
to the brane part of the solution will coincide with the metric of
a global monopole of Ref. \cite{Vil}.

The exterior region outside the outer horizon at ${r_H}_+$ interpolates between
the near horizon region and the asymptotic region described by metrics
(\ref{nbbas}) and (\ref{bnhm}), respectively. In the far region, the spacetime
described by the metric (\ref{nbbas}) asymptotes to $AdS_D$, even though the
longitudinal part of the solution is not flat. The Riemann curvature tensor
corresponding to the metric (\ref{nbbas}) is
\begin{equation}
R_{MNPQ}=\left(-B_\infty+\left[\frac{H^2}{B_\infty r^2}
\right]_{M,N,P,Q\in\{x^\mu\}}+\left[\frac{1}{r^2}\right]_{M,N,P,Q
\in\{\theta^a\}}\right)(g_{MP}g_{NQ}-g_{MQ}g_{NP}),
\end{equation}
where the two terms dressed by square brackets are non-vanishing
only if $M,N,P,Q\in\{x^\mu\}$ and $M,N,P,Q\in\{\theta^a\}$,
respectively. Since the two terms in the square bracket decay
quadratically, the spacetime asymptotes to $AdS_D$ as
$r\to\infty$. This type of spacetime has also been found in
Ref. \cite{Vil}, and the metric can be rewritten to the form found
in Ref. \cite{Vil} as done in the flat brane case
\begin{equation}
ds^2\approx e^{2\sqrt{B_\infty}\xi}\hat{g}_{\mu\nu}(x)dx^\mu dx^\nu+d\xi^2
+e^{2\sqrt{B_\infty}\xi}d\Omega_{d-1}^2.
\end{equation}

\begin{figure}[htbp]
\centerline{\epsfig{file=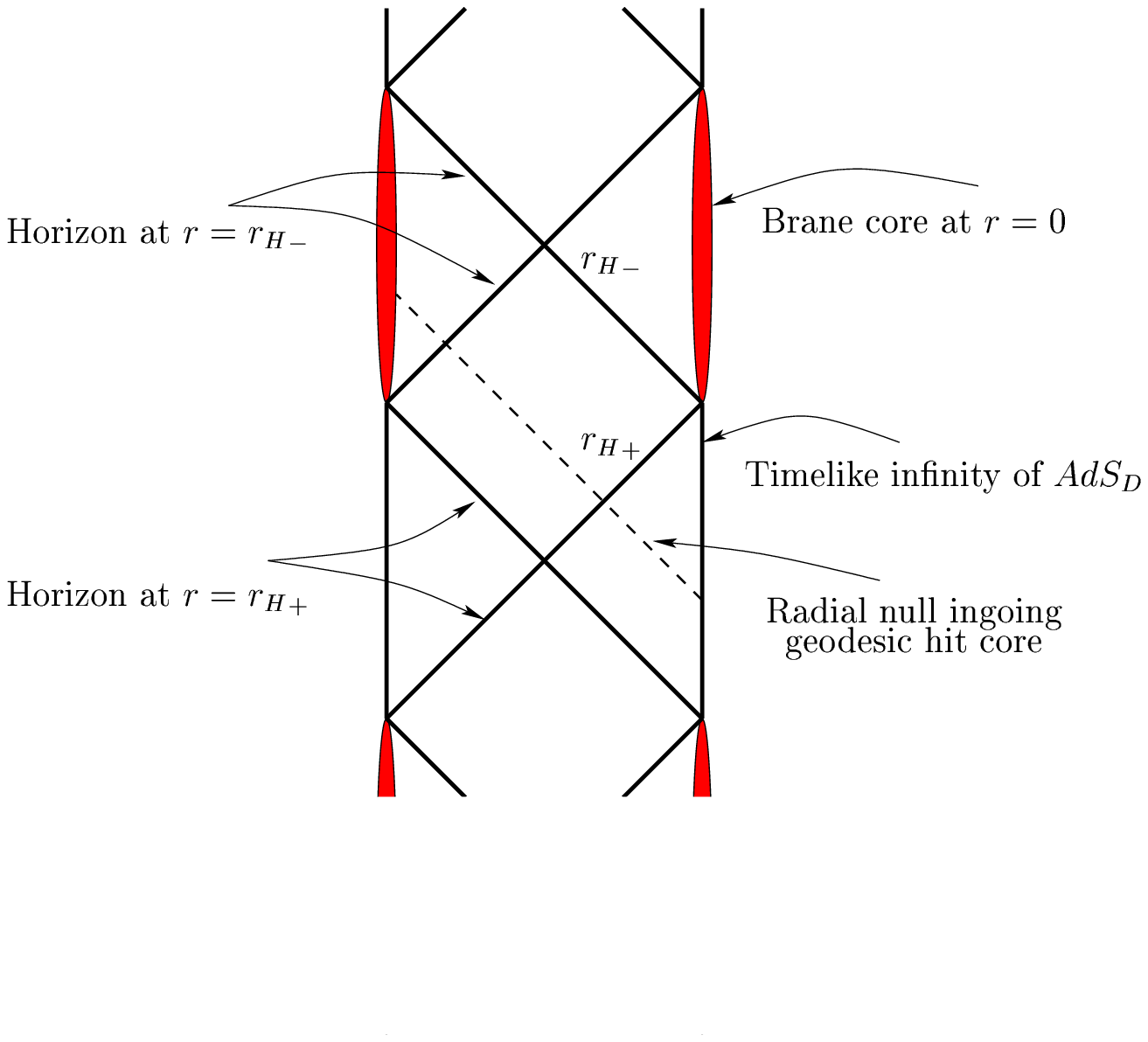, width=10cm}} \caption{
Penrose diagram for the non-extreme global black brane with field
configurations shown in FIG. \ref{bbb}. The spatial coordinates
$x^i$ and $\theta^a$ are suppressed. This diagram is very similar
to that of the non-extreme Reissner-Nordstr\"{o}m black hole, but
now there is a non-singular brane core at $r=0$ rather than a
curvature singularity. The boundary at $r=\infty$ is timelike
rather than null.} \label{nebbs}
\end{figure}

Now we turn to the near horizon metrics (\ref{bnhm}). One may check directly
by calculating the geodesics of the metric (\ref{bnhm}) that the coordinate
chart $\{x^\mu,r,\theta^a\}$ is geodesically incomplete.
As done in the case of flat brane, we consider the motion of test particles.
It will be sufficient to consider the radial motion of test particles along
the line of $\gamma=0$ in the static coordinate system to see the causal
structure on the transverse slice. That is, it is sufficient to study
null and timelike geodesics in the system
\begin{equation}\label{trmet}
ds^2=-\frac{4H^2}{{B_1}_\pm}(r-{r_H}_\pm)dt^2
+\frac{dr^2}{{B_1}_\pm(r-{r_H}_\pm)},
\end{equation}
where we have used the relation (\ref{HH}). This metric is
reminiscent of the near horizon metrics of the non-extremal
Reissner-Nordstr\"{o}m black hole. The proper distances from a
point $r_0$ to the horizons in a constant time slice is finite and
is given by $\chi(r_0)\sim2\sqrt{(r_0-{r_H}_\pm)/{B_1}_\pm}$. With
this metric we find the geodesic equation
\begin{equation}
\kappa=-\frac{H^2}{{B_1}_\pm}(r-{r_H}_\pm)\left(\frac{dt}{d\tau}\right)^2
+\frac{1}{{B_1}_\pm(r-{r_H}_\pm)}\left(\frac{dr}{d\tau}\right)^2.
\end{equation}
With the static world volume metric, there is a constant of motion
corresponding to the Killing vector $\partial_t$
\begin{equation}
\epsilon=\frac{H^2}{{B_1}_\pm}(r-{r_H}_\pm)\frac{dt}{d\tau}.
\end{equation}
The equation of motion of the test particles then is
\begin{equation}
\left(\frac{dr}{d\tau}\right)^2-\epsilon^2\frac{{B_1}_\pm}{H^2}
-\kappa {B_1}_\pm(r-{r_H}_\pm)=0.
\end{equation}
The equation of motion tells that the test particle reach at the horizon with
a finite velocity, i.e., $(dr/d\tau)\to\epsilon {B_1}_\pm/H$ as
$r\to{r_H}_\pm$. Therefore, the coordinate chart $\{x^\mu,r,\theta^a\}$ is
geodesically incomplete and so the extensions beyond each horizon are required.
If the solutions obtained in separated regions are mutually connected as shown
in FIG. \ref{bbb}, the causal structure induced by the $(t,r)$-part of the
solutions will resemble that of the non-extreme RN black hole.
The differences are the non-singular brane core instead a curvature singularity
at $r=0$ and the timelike boundary rather than null at $r=\infty$.
The corresponding Penrose diagram is shown in FIG. \ref{nebbs}.

The metric (\ref{bnhm}) has some very interesting properties. The metrics
with the form of (\ref{bnhm}) usually have curvature singularities at
$r={r_H}_\pm$. However, when the relation (\ref{HH}) holds, the curvature
invariants are finite and the surfaces at $r={r_H}_\pm$ do indeed appear
to be just coordinate singularities. For example, the corresponding
Kretschmann scalar near each horizon at $r={r_H}_\pm$ is given by
\begin{equation}
R_{MNPQ}R^{MNPQ}=\frac{2(d-1)(d-2)}{{r_H}_\pm^4}.
\end{equation}
This is not surprising, since the metric (\ref{bnhm}), in fact, describes
constant curvature spacetimes with topology of locally $(p+2)$-dimensional
flat Minkowski spacetime multiplied by $(d-1)$-dimensional sphere with
radius ${r_H}_\pm$, i.e., ${\cal M}^{1,p+1}\times S^{d-1}$. This can
easily be observed by calculating the Riemann curvature tensor and noticing
that the only non-vanishing components of the corresponding Riemann
curvature tensor are
\begin{eqnarray}
R_{\theta_a\theta_b\theta_c\theta_d}=\frac{1}{{r_H}_\pm^2}(g_{\theta_a\theta_c}
g_{\theta_b\theta_d}-g_{\theta_a\theta_d}g_{\theta_b\theta_c}),
\end{eqnarray}
where $\theta_a,\theta_b,...$ refer to indices on $S^{d-1}$.

It is interesting to observe that the $(p+2)$-dimensional flat part corresponds
the spacetime of a $p$-dimensional thin domain wall embedded in
$(p+2)$-dimensional flat spacetime, which is a generalization
of the inflating three-brane in four-dimensions \cite{vilenkin}.
The metric (\ref{bnhm}) can be rewritten to an analogous form to that of the
inflating three-brane found in Ref. \cite{vilenkin} introducing a new radial
coordinate $e^{-2Hz}\equiv 4H^2[(r-{r_H}_\pm)/{B_1}_\pm]$, the metric can
easily be rewritten as
\begin{equation}\label{dbr2}
ds^2=e^{-2Hz}\left[d\hat{s}^2+dz^2\right]+{r_H}_\pm^2d\Omega_{d-1}^2,
\end{equation}
where the horizons are located at $z=\infty$. Another coordinate transformation
$\rho\equiv2\sqrt{{B_1}_\pm(r-{r_H}_\pm)}/{B_1}_\pm+\rho_0$, where $\rho_0$
is an integral constant, brings Eq. (\ref{bnhm}) to the form
\begin{equation}\label{dbr}
ds^2=(1-H\rho)^2d\hat{s}^2+d\rho^2+{r_H}_\pm^2d\Omega_{d-1}^2,
\end{equation}
where the relation Eq. (\ref{HH}) has been used and $\rho_0$ has
been set as to be $\rho_0=H^{-1}$, so that the horizons are at
$\rho=H^{-1}$. The $(x_\mu,\rho)$-parts of metrics explicitly
coincide with those of the vacuum domain wall obtained in
Ref. \cite{vilenkin}. Thus, the near horizon spacetime of the bent brane
is nothing but that of $(p+2)$-dimensional vacuum domain wall times
$(d-1)$-dimensional sphere.

The spacetime of domain walls has been extensively studied in
many literatures \cite{vilenkin,widrow,gibbons,WL,WL2,CGR,CDGS,CGS,kl,kaloper}.
For the metric Eq. (\ref{dbr2}), the $z={\rm const}$ (or $\rho={\rm const}$
for Eq. (\ref{dbr})) hypersurfaces have the properties of
$(p+1)$-dimensional de Sitter space. Furthermore, the
$(t,z)$-part (or $(t,\rho)$-part) of the metric describes a
$(1+1)$-dimensional Rindler space (i.e., flat space in the
reference frame of a uniformly accelerating observer with proper
acceleration $H$). Therefore, at each of the $(p+1)$ spatial
directions event horizons exist. The ones in the
$p$ longitudinal directions are de Sitter-like as discussed at the
beginning of this subsection. The one in the radial direction is
Rindler-like, and the various extensions beyond it were considered
in Refs. \cite{vilenkin,gibbons,WL2}. The extensions could be
simply obtained by gluing together two portions of Minkowski
spacetime. Such extensions occupy only the interior region of the
Penrose diagram of FIG. \ref{nebbs} without the exterior region
interpolating between the horizon and the timelike infinity.
On the other hand, the global spacetime of non-extreme global
black brane corresponding to the field configuration of FIG. \ref{bbb}
bears remarkable similarities to that of the non-extreme domain
walls found in Ref. \cite{CGS}.

When the horizon size ${r_H}_-$ is of the order of the Planck
length scale (this will be true if $|\Lambda|^{1/2}\sim M_*\sim
M_{pl}$), at low-energy below the Planck scale the brane core will
look like a inflating domain wall embedded in $(p+2)$-dimensional
Minkowski spacetime to an observer residing near the inner
horizon. However, when the bulk cosmological constant is much less than
the fundamental scale, the horizon size is much larger than the
fundamental length scale and so, even at low-energy, the core will
look like a string, or a monopole, etc, depending on the number of
extra dimensions rather than a domain wall. Interestingly, when
$p=1$ and $d=2$ the metric (\ref{dbr}) itself coincide with the
asymptotic solution near the horizon of the non-static global
cosmic string found in Ref. \cite{nsgs}, and the interior
region inside the inner horizon resembles the spacetime of the
non-static global cosmic string. On the other hand, the nature of
the cosmological event horizon of Refs. \cite{nsgs,WN} seems to be
different from the inner horizon at ${r_H}_-$, in the sense that
across the cosmological event horizon the coordinates $t$ and $\rho$ of
\cite{nsgs,WN} remain timelike and spacelike respectively, while
the roles of $t$ and $\rho$ in the metric (\ref{dbr}) is exchanged
across the inner horizon.\footnote{ This can be seen explicitly from
the metric (\ref{bnhm}). Note that, if ${r_H}_-<r<{r_H}_+$,
the coordinate transformation from $r$ to $\rho$ should be replaced by
$\rho\equiv-2\sqrt{{B_1}_\pm(r-{r_H}_\pm)}/{B_1}_\pm+\rho_0$ and
the metric is replaced by
$ds^2=-(1-H\rho)^2d\hat{s}^2-d\rho^2+{r_H}_\pm^2d\Omega_{d-1}^2$.}
The interior region of the other defects of $p>1$ and $d\geq2$ are
simply the generalizations of the non-static global cosmic string
spacetime.

\section{Brane worlds and a thermal instability for the cosmological
constant}\label{secBWM}
~\\
In this section we consider the global black branes as an approximation
for brane world scenarios. We will shortly summarize the contents
of Ref. \cite{KMR}, in which the extremal ones only are considered, and
discuss the effective gravity on the brane when it is non-extremal.
We briefly comment a few thermodynamic properties of the global black
branes, and discuss a decrease of the effective cosmological constant
in the brane world through thermal instability of the non-extremal
black brane. The exterior region of the extremal global black branes
by itself does not seem to provide a suitable
framework for the brane world scenario, because in one direction
one will reach the $(p+2)$-dimensional AdS horizon, while in the
other direction one will reach the $D$-dimensional AdS spacetime.
On the other hand, the interior region possesses all the feature
necessary for realizing the Randall-Sundrum (RS) type brane world
scenario. In one direction, it asymptotes to the infinitely long
AdS throat. As such, the central region surrounded by the AdS
region looks like a one-sided RS domain-wall embedded
in $(AdS_{p+2})$. As discussed in Ref. \cite{KMR}, the size of the
horizon can be interpreted as the effective size of $d$ extra
dimensions in that the Planck scale in the $p$-dimensional brane
world $M_{pl}$ is determined by the fundamental scale $M_*$ and
the horizon size $r_H$ via the familiar relation $M_{pl}^{p-1}\sim
M_*^{p-1+d}r_H^d$. Moreover, the gravity on the brane world
behaves as expected in a world with $d$-extra dimensions
compactified with size $r_H$.

When the fundamental scale and the bulk cosmological constant are
of the order of the brane world Planck scale, i.e.,
$M_*\sim|\Lambda|^{1/2}\sim M_{pl}$, then the curvature radii of
$AdS_{p+2}$ and $S^{d-1}$ are of the order of the Planck scale.
Thus, at low-energy below the Planck scale, the extra space is
reduced effectively to a one-dimensional space. Consequently, the
global $p$-brane core looks like a $p$-dimensional domain-wall
embedded in an $AdS_{p+2}$ bulk spacetime. In the \lq thin
core-approximation' limit, the physics on the brane is essentially
the same as that of the original RS scenario. On the
other hand, when the cosmological constant is hierarchically
smaller than the fundamental mass scale, the horizon size is much
larger than the core size or the fundamental length scale, i.e.,
$r_H\gg r_c\sim M_*^{-1}$, so the physics is quit analogous to
that of the large extra dimension scenario. Then the Planck scale
$M_{pl}$ in the brane world is hierarchically bigger than the
fundamental scale $M_*$ of the higher dimensional gravity,
according to the familiar relation $M_{pl}^{p-1}\sim
M_*^{p-1+d}r_H^d$. And as shown in Ref. \cite{KMR}, the
phenomenology on the brane is imperceptibly different from that of
the usual large extra dimension scenario.

For the bent branes surrounded by two horizons, since the interior region
is bounded by the inner horizon as the interior region of the extreme
black brane is bounded by the Cauchy horizon, the outside of the
inner horizon is out of causal contact with the brane world. So
the Planck scale on the brane world is given by $M_{pl}^{p-1}\sim
M_*^{p-1+d}{r_H}_-^d$. The interior region has similar features to
the inflating domain wall spacetime or the de Sitter brane, so the
gravity on the brane will resemble that on the de Sitter brane. As
discussed in Refs. \cite{gubser,karch,garriga}, the massless
graviton is trapped on the de Sitter brane reproducing the correct
$(p+1)$-dimensional gravity on the brane as on the Minkowski
brane. A difference is that in the Minkowski case the massless
graviton is a marginal bound state and the continuum of
Kaluza-Klein modes starts at zero mass, while in the de Sitter
case the graviton is separated by a finite mass gap of $m=(3/2)H$
from the continuum, that is, the Kaluza-Klein continuum modes
starts at $m=(3/2)H$.

In this picture, the Hawking radiation could be a possible mechanism
for the resolution of various problems associated with the brane
world scenarios, such as the observed flatness and the approximate
Lorentz invariance of our world. The bending of the brane and the bulk
curvature that violate the $SO(3,1)$ isometry on our brane
\cite{kolb,poritz}
would correspond to excitations upon the extremal state,\footnote{
Here, the corresponding non-extremal black branes are different from those
discussed in this paper. Rather, they will resemble the static black branes,
of which world volume metrics depend on the extra dimension coordinates,
e.g., $d\hat{s}^2=\hat{g}_{\mu\nu}(y)dx^\mu dx^\nu=-dt^2+h(r)dx_i^2$.}
and the excited states that correspond to non-extremal states would
evolve into an extremal state through the Hawking radiation process.
The similar argument may be applicable as well for the cosmological
constant problems.

We now turn to the thermal instability of the non-extremal black
branes, which could be a mechanism resolving the cosmological constant
problem.\footnote{There have been many interesting ideas in the context
of the brane world scenario in attempts to solve the cosmological
constant problem \cite{ADKS,KMS,KT,tetra,KKL}. The basic idea
is \lq self tuning'. The brane tension could take on an extended range of
values without jeopardizing the stability of the cosmological constant
on the brane. In this way, the brane could be stable against any radiative
corrections to the brane tension. However, the naked singularities, which
lead to inconsistencies in the effective theory on the brane world, is
inherent \cite{FLLN,CEGH}, or the static self-tuned solutions are
dynamically unstable \cite{LZ,BCG,DMW,Med}.} In the brane world,
the intrinsic curvature of the brane amounts to a net cosmological
constant as viewed by an observer pinned on the brane, i.e.,
$\Lambda_{\rm phys}=[(p-1)/2(p+1)]\hat{R}$. Since we have found
that, in higher-dimensional pure gravity theory with a
cosmological constant term $\Lambda$ explicitly, there exists a
class of solutions with physical $(p+1)$-dimensional metrics
obeying the standard Einstein equations with positive arbitrary
(including zero) values of $\Lambda_{\rm phys}$ or $\hat{R}$, the
cosmological constant problem may be solved by choosing a solution
characterized by vanishingly small $\Lambda_{\rm phys}$, that is,
a very near extremal black brane solution.\footnote{Recent cosmological
observations \cite{BOPS} place the cosmological constant at a very
small value (but not zero) in comparison to the Planck scale; in fact,
$\bar{\Lambda}_{\rm phys}\equiv M_{\rm Pl}^2\Lambda_{\rm phys}/8\pi
\approx10^{-120}M_{\rm Pl}^4$.}
However, this doesn't seem to tell
anything about the cosmological constant problem, because there
are no convincing arguments in favor of this choice. In this
respect, we might need a dynamical mechanism that singles out the
solution with an extremely small $\Lambda_{\rm phys}$, which would be a
complete solution to the cosmological constant problem. As discussed in
Refs. \cite{KMR,BCM}, a possible dynamical mechanism may be related
to the thermodynamic instability of the non-extreme black branes,
by which the positive vacuum energy density on the brane could be
radiated away causing the brane to evolve into an extremal state.

In this respect, we briefly comment a few thermodynamic properties
of the extreme and non-extreme global black branes, and discuss a
decrease of the effective cosmological constant on the brane. For
the extremal black brane, since the surface gravity is zero at the
degenerated horizon, the Hawking temperature is zero and there is
no Hawking radiation. Thus, this system is in fact
thermodynamically stable. The entropy of a system is associated
with the number of accessible states. Since the extremal state of
the global black brane seems to be a unique solution to the
equations of motion as discussed in the subsection \ref{secSolsA}, its
entropy would be zero. On the other hand, the entropy associated
with a spacetime containing horizon is given by the area of the
event horizon:
$S_{BH}=A_D/4\pi G_D$,
where $A_D$ is the area of the horizon and $G_D$ is the $D$-dimensional
Newton constant. The area of the spacelike surface of constant $r$ reads from
the metric Eq.(\ref{bbhm}) as:
$A_D(r)\sim (r-r_H)^{2p(1-\alpha)}$,
which clearly vanishes as $r\to r_H$.
This supports the conclusion of zero entropy property of the extreme black
brane.

For the non-extremal case, as mentioned in the previous section
there are de Sitter horizon in $p$ longitudinal directions and the
Rindler horizon in radial direction. The de Sitter horizon is
associated with a surface gravity $\kappa=H$ and so an observer
will detect an isotropic background in the longitudinal directions
of thermal radiation with temperature $T=(2\pi)^{-1}H$ coming,
apparently, from the horizon \cite{GH}. The de Sitter horizon also
may be interpreted as the entropy or lack of information that the
observer has about the regions of the universe that he cannot see.
If one absorbs the thermal radiations, one gains energy and entropy
at the expanse of this region and so the area of the horizon will
go down according to the first law of the black hole dynamics.
As the area decreases, the temperature of the cosmological radiation
goes up (like the black hole case), so the cosmological event
horizon is unstable.
On the other hand, at the Rindler horizon the surface gravity can
be identified by the proper acceleration of a uniformly
accelerating observer. Since the metric corresponds to that of the
uniformly accelerating observer with the proper acceleration $H$,
the surface gravity is simply given by $\kappa=H$, and the
temperature measured by a static observer near both horizons at
${r_H}_\pm$ is just $T=(2\pi)^{-1}H$ or
$T=(2\pi)^{-1}\sqrt{2\Lambda_{\rm phys}/p(p-1)}$. Therefore, the
observer will detect two thermal radiations with the same
temperatures of $T=(2\pi)^{-1}H$ coming from the de Sitter horizon
and the Rindler one, respectively. This seems to be well matched
with the zeroth law of black hole mechanics. Here, an interesting
point is that the surface gravity at the horizons is determined by the
brane world volume curvature solely regardless the size of the
horizon unlike usual black hole and static black brane cases.
An observer residing in the exterior region at a distance from the
outer horizon may observe a steady current of thermal radiation coming
from the outer horizon and going out toward the asymptotically anti-de
Sitter infinity, even if the observer see an isotropic thermal background
in the longitudinal directions. This implies a possibility that the
non-extreme black brane could evolve into an extreme black brane of
which world volume is flat. Since the brane itself is the source
of the radiation in radial direction, this must result in energy loss
from the brane. If the quantum fluctuations of the vacuum energy on
the brane is coupled to bulk fields (including bulk gravitational
field), then the vacuum energy density could be radiated into the
bulk outside of the outer horizon through the Hawking radiation
process, resulting in a decrease of the vacuum energy density or,
equivalently, the world volume curvature $\hat{R}$. Thus, the Hawking
temperature goes down with time, and the radiation will last until
the temperature reaches to zero. Since the temperature is proportional
to $\sqrt{\hat{R}}$, the radiation will go on until $\hat{R}=0$.
Equivaletly, since the temperature is proportional to the square root
of the effective cosmological constant, $\sqrt{\Lambda_{\rm phys}}$,
the radiation will continue until $\Lambda_{\rm phys}=0$.
Thus, this effect may radiate away all contributions to the cosmological
constant.\footnote{Recently, a very interesting proposal on a thermal
instability for the cosmological constant was appeared in Ref.
\cite{smolin}. There are modes of the linearized bulk gravitational
field which see the brane as a accelerating mirror. This give rise to
an emission of thermal radiation from the brane into the bulk.
The temperature is also proportional to the square root of the
cosmological constant on the brane world, $\sqrt{\Lambda_{\rm phys}}$.}

\section{Conclusions}\label{secSD}
~\\
In this paper we have mainly examined global black $p$-brane solutions,
which are black hole-like $p$-dimensional extended global defect
solutions to a scalar theory with global $O(d)$ internal symmetry
coupled to higher-dimensional gravity with a negative cosmological
constant. The spacetimes of them are perfectly regular everywhere.
We have found series solutions in a few separated regions and
numerically showed that they can be smoothly connected. There are
two kinds of such solutions, extremal or non-extremal black branes.
The extremal black branes have a Ricci flat world volume and are
surrounded by a degenerated Killing horizon. On the other hand,
the world volumes of non-extremal black branes surrounded by two
horizons are not Ricci flat but of non-zero constant curvature.
The extremal black brane solutions seem to be the critical limit
of non-extremal ones, in that the positions of two horizons of
non-extremal branes approach to and are degenerated at the position
the horizon of extremal black branes in the limit that the world
volume curvature goes to zero, i.e., $\hat{R}\to0$. The exterior
region of both objects asymptotes to anti-de Sitter infinity.

The near horizon geometries of both the extremal and the non-extremal
branes have very interesting features. For the extremal black
brane the geometry is the infinitely long anti-de Sitter throat
with the topology of $AdS_{p+2}\times S^{d-1}$. The curvature
scales of both $AdS_{p+2}$ and $S^{d-1}$ parts are of the order of
$\sqrt{|\Lambda|}$. So the core will look like a one-side
flat domain wall embedded in $(p+2)$-dimensional anti-de Sitter
space to an observer residing in the near horizon region of the
interior region if the size of the sphere is small so that
the observer doesn't see $S^{d-1}$-part. The physics on the branes
then is essentially the same as that on the Randall-Sundrum brane
as discussed in Ref. \cite{KMR}. On the other hand, for
the non-extremal ones it is the $(p+2)$-dimensional flat spacetime
times $(d-1)$-dimensional sphere with radii ${r_H}_\pm$ at each
horizon. The $(p+2)$-dimensional flat part corresponds to the
spacetime of a $p$-dimensional thin domain wall embedded in
$(p+2)$-dimensional flat spacetime. An observer residing in the
near horizon region inside the inner horizon will see an inflating
domain wall if the size of the inner horizon is small enough.
Therefore, the gravitational Kaluza-Klein modes have a mass gap
in the spectrum and the continuous spectrum starts above the mass
$m=(3/2)H$, where $H$ is the expansion rate of the brane.

In the brane world, the world volume curvature amounts to the
effective cosmological constant. We obtained a flat brane
solution with $\hat{R}=0$ (i.e., the extremal black brane) without
fine-tuning between parameters in the Lagrangian. However, to solve the
cosmological constant problem completely we need a dynamical mechanism that
picks out the flat brane solution among whole solutions of any $\hat{R}$.
We have suggested a possible thermodynamic mechanism as a candidate
for such mechanism. It is provable for a non-extremal black brane
to evolve into an extremal black brane owing to its thermodynamic
instability. It might emit a thermal radiation with temperature
$T\propto\sqrt{\hat{R}}\propto\sqrt{\Lambda_{\rm phys}}$ resulting
in a decrease of the vacuum energy on the brane, flattening itself
and finally reaching to the extremal limit at temperature $T=0$.
This effect will dilute all contribution to the cosmological constant.
If it is really possible, then the black brane world model represents
a solution to the cosmological constant problem.

\section*{Acknowledgements}
~\\
We would like to thank Yoonbai Kim and Soo-Jong Rey for collaboration on
the previous work and for useful discussions. We also would like to thank
Hyeong-Chan Kim for delightful discussions.
It is a great pleasure to thank Yong-Jin Chun for his assistance to the
numerical works. Without his aid, it would have not been finished.
This work was supported by the BK21 project of Ministry of Education.




\begin{thebibliography}{99}

\bibitem{VS} A. Vilenkin and E.P.S. Shellard, {\it Cosmic String and other
Topological Defects}, Cambridge University Press, 1994.
\bibitem{gss} R. Gregory, Phys. Lett. {\bf B 215}, 663 (1988);
              G.W. Gibbons, M.E. Ortiz and F. Ruiz-Ruiz, Phys. Rev.
              {\bf D 39}, 1546 (1989).
\bibitem{CK1} A. G. Cohen and D. B. Kaplan, Phys. Lett. {\bf B 215}, 67 (1988).
\bibitem{BV} M. Barriola and A. Vilenkin, Phys. Rev. Lett. {\bf 63},
341 (1989).
\bibitem{nsgs} R. Gregory, Phys. Rev. {\bf D 54}, 4955 (1996).
\bibitem{WN} A. Wang and J.A.C. Nogales, Phys. Rev. {\bf D 56}, 6217 (1997).
\bibitem{ADD} N. Arkani-Hamed, S. Dimopoulos and G. Dvali, Phys. Lett.
       {\bf B429} (1998) 263; Phys. Rev. {\bf D 59} (1999) 086004;
\bibitem{RS1} L. Randall and R. Sundrum, Phys. Rev. Lett. {\bf 83}, 3370 (1999).
\bibitem{RS2} L. Randall and R. Sundrum, Phys. Rev. Lett. {\bf 83}, 4690 (1999).
\bibitem{CK2} A. G. Cohen and D. B. Kaplan, Phys. Lett. {\bf B 470}, 52 (1999).
\bibitem{Gre} R. Gregory, Phys. Rev. Lett. {\bf 84}, 2564 (2000).
\bibitem{Vil} I. Olasagati and A. Vilenkin, Phys. Rev. {\bf D 62},
044014 (2000).
\bibitem{shap1} T. Gherghetta and M. Shaposhnikov, Phys. Rev. Lett.
{\bf 85}, 240 (2000).
\bibitem{shap} T. Gherghetta, E. Roessl and M. Shaposhnikov,
              Phys. Lett. {\bf B 491}, 353 (2000).
\bibitem{Oda} I. Oda, Phys.Lett. {\bf B 496}, 113 (2000).
\bibitem{KMR} Y. Kim, S.-H. Moon and S.-J. Rey, Nucl. Phys. B {\bf 602}, 467
(2001).
\bibitem{inyong} K. Benson and I. Cho, Phys. Rev. {\bf D 64}, 065026 (2001).
\bibitem{BCM} W.S. Bae, Y.M. Cho and S.-H. Moon, J. High-Energy Phys.
{\bf 0103}, 039 (2001).
\bibitem{KKK} N. Kim, Y. Kim and K. Kimm, Phys. Rev. {\bf D 56}, 8029 (1997);
Class. Quant. Grav. {\bf 15}, 1513 (1998).
\bibitem{KM} Y. Kim and S.-H. Moon, Phys. Rev. {\bf D 58}, 105013 (1998);
G. Cl\'{e}ment and A. Fabbri, Class. Quant. Grav. {\bf 17}, 2537 (2000).
\bibitem{LNW} K. Lee, V.P. Nair and E.J. Weinberg, Phys. Rev. Lett. {\bf 68},
1100 (1992);  Phys. Rev. {\bf D 45}, 2751 (1992).
\bibitem{HS} G.T. Horowitz and A. Strominger,
            Nucl. Phys. B {\bf 360}, 197 (1991).
\bibitem{vilenkin} A. Vilenkin, Phys. Lett. {\bf B 133}, 177 (1983).
\bibitem{widrow} L.M. Widrow, Phys. Rev. {\bf D 39}, 3571 (1989).
\bibitem{gibbons} G.W. Gibbons, Nucl. Phys. B {\bf 394}, 3 (1993).
\bibitem{WL} A. Wang and P.S. Letelier, Phys. Rev. {\bf D 51}, R6612 (1995).
\bibitem{WL2} A. Wang and P.S. Letelier, Phys. Rev. {\bf D 52}, 1800 (1995).
\bibitem{CGR} M. Cvetic, S. Griffies and S.-J. Rey, Nucl. Phys. B {\bf 381},
301 (1992); Nucl. Phys. B {\bf 389}, 3 (1993).
\bibitem{CDGS} M. Cvetic, R.L. Davis, S. Griffies, and H.H. Soleng,
Phys. Rev. Lett. {\bf 70}, 1191 (1993).
\bibitem{CGS} M. Cvetic, S. Griffies, and H.H. Soleng,
Phys. Rev. Lett. {\bf 71}, 670 (1993).
\bibitem{kl} N. Kaloper and A. Linde, Phys. Rev. {\bf D 59}, 101303 (1999).
\bibitem{kaloper} N. Kaloper, Phys. Rev. {\bf D 60}, 123506 (1999).
\bibitem{GH} G.W. Gibbons and S.W. Hawking, Phys. Rev. {\bf D 15}, 2738 (1977).
\bibitem{gubser} O. DeWolfe, D.Z. Freedman, S.S. Gubser, and A. Karch,
Phys. Rev. {\bf D 62}, 046008 (2000).
\bibitem{karch} A. Karch and L. Randall, J. High-Energy Phys.
{\bf 0105}, 008 (2001).
\bibitem{garriga} J. Garriga and M. Sasaki,
Phys. Rev. {\bf D 62}, 043523 (2000).
\bibitem{kolb} D.J.H. Chung, E.W. Kolb, and A. Riotto, {\it Extra dimensions
present a new flatness problem}, hep-ph/0008126.
\bibitem{poritz} J.F. V\'{a}zquez-Poritz, {\it Massive Gravity on a
Non-extremal Brane}, hep-th/0110299.
\bibitem{rubakov} V.A. Rubakov and M.E. Shaposhnikov,
Phys. Lett. B {\bf 125} (1983) 139.
\bibitem{ADKS} N. Arkani-Hamed, S. Dimopoulos, N. Kaloper, and R. Sundrum,
Phys. Lett. B {\bf 480}, 193 (2000).
\bibitem{KMS} S. Kachru, M. Schulz, and E. Silverstein,
Phys. Rev. {\bf D 62}, 045021 (2000).
\bibitem{KT} A. Kenhagias and K. Tamvaski, {\it A Self-Tuning Solution
of the Cosmological Constant Problem}, hep-th/0011006.
\bibitem{tetra} N. Tetradis, Phys. Lett. B {\bf 509}, 307 (2001).
\bibitem{KKL} J.E. Kim, B. Kyae and H.M. Lee,
Phys. Rev. Lett. {\bf 86}, 4223 (2001); Nucl. Phys. B {\bf 613}, 306 (2000).
\bibitem{FLLN} S. Forste, Z. Lalak, S. Lavignac, and H.P. Nilles,
Phys. Lett. B {\bf 481}, 360 (2000); J. High-Energy Phys. {\bf 0009}, 034 (2000).
\bibitem{CEGH} C. Csaki, J. Erlich, C. Grojean, and T. Hollowood,
Nucl. Phys. B {\bf 584}, 359 (2000).
\bibitem{LZ} I. Low and A. Zee, Nucl. Phys. B {\bf 585}, 395 (2000).
\bibitem{BCG} P. Binetruy, J.M. Cline and C. Grojean,
Phys. Lett. B {\bf 489}, 403 (2000).
\bibitem{DMW} J. Diemand, C. Mathys and D. Wyler, {\it Dynamical
Instabilities of Brane World Models}, hep-th/0105240.
\bibitem{Med} A.J.M. Medved, {\it Dynamical Instability of Self-Tuning
Solution with Antisymmetric Tensor Field}, hep-th/0109180.
\bibitem{BOPS} N. Bahcall, J.P. Ostriker, S. Perlmutter and P.J. Steinhardt,
Science {\bf 284}, 1481 (1999).
\bibitem{smolin} S. Alexander, Yi Ling, and Lee Smolin, {\it A thermal
instability for positive brane cosmological constant in the Randall-Sundrum
cosmologies}, hep-th/0106097.

\end{thebibliography}
\end{document}